\documentclass[%
 reprint,
 amsmath,amssymb,
 aps,longbibliography,superscriptaddress
]{revtex4-2}

\usepackage{graphicx}
\usepackage{dcolumn}
\usepackage{bm}
\usepackage{multirow,stackengine}
\usepackage{hyperref}
\hypersetup{pdfencoding=auto,colorlinks=true,linkcolor=blue,citecolor=blue}

\begin{document}

\preprint{APS/123-QED}

\title{Electronic transport in double-nanowire superconducting islands with multiple terminals}

 \author{Alexandros Vekris}
 \affiliation{Center for Quantum Devices, Niels Bohr Institute, University of Copenhagen, 2100 Copenhagen, Denmark}
 \affiliation{Sino-Danish Center for Education and Research (SDC) SDC Building, Yanqihu Campus, University of Chinese Academy of Sciences, 380 Huaibeizhuang, Huairou District, 101408 Beijing, China}

 \author{Juan Carlos Estrada Salda\~na} 
 \affiliation{Center for Quantum Devices, Niels Bohr Institute, University of Copenhagen, 2100 Copenhagen, Denmark}
  
 \author{Thomas Kanne}
 \affiliation{Center for Quantum Devices, Niels Bohr Institute, University of Copenhagen, 2100 Copenhagen, Denmark}

 \author{Thor Hvid-Olsen} 
 \affiliation{Center for Quantum Devices, Niels Bohr Institute, University of Copenhagen, 2100 Copenhagen, Denmark}

\author{Mikelis Marnauza}
 \affiliation{Center for Quantum Devices, Niels Bohr Institute, University of Copenhagen, 2100 Copenhagen, Denmark}

 \author{Dags Olsteins}
 \affiliation{Center for Quantum Devices, Niels Bohr Institute, University of Copenhagen, 2100 Copenhagen, Denmark}

\author{Matteo M. Wauters}
\affiliation{Center for Quantum Devices, Niels Bohr Institute, University of Copenhagen, 2100 Copenhagen, Denmark}
\affiliation{Niels Bohr International Academy, Niels Bohr Institute, University of Copenhagen, 2100 Copenhagen, Denmark}

\author{Michele Burrello}
\affiliation{Center for Quantum Devices, Niels Bohr Institute, University of Copenhagen, 2100 Copenhagen, Denmark}
\affiliation{Niels Bohr International Academy, Niels Bohr Institute, University of Copenhagen, 2100 Copenhagen, Denmark}

\author{Jesper Nyg{\aa}rd}
\affiliation{Center for Quantum Devices, Niels Bohr Institute, University of Copenhagen, 2100 Copenhagen, Denmark}
    
\author{Kasper Grove-Rasmussen}
\affiliation{Center for Quantum Devices, Niels Bohr Institute, University of Copenhagen, 2100 Copenhagen, Denmark}

\date{\today}

\begin{abstract}
We characterize in-situ grown parallel nanowires bridged by a superconducting island. The magnetic-field and temperature dependence of Coulomb blockade peaks measured across different pairs of nanowire ends are consistent with a sub-gap state extended over the hybrid parallel-nanowire island. Being gate-tunable, accessible by multiple terminals and free of quasiparticle poisoning, these nanowires show promise for the implementation of several proposals that rely on parallel nanowire platforms.

\end{abstract}

\maketitle


\section{Introduction}

In-situ grown double nanowires are at the center of research regarding qubit devices~\cite{Aasen2016Aug}, coupled sub-gap states~\cite{Yao2014Dec,Kurtossy2021Oct} and exotic topological superconductivity, as a plethora of theoretical proposals~\cite{GaidamauskasPRL2014,Klinovaja2014Jul,EbisuProg2016,SchradePRB2017,ReegPRB2017,SchradePRL2018,ThakurathiPRB2018,DmytrukPRB2019,ThakurathiPRR2020,KotetesPRL2019,PapajPRB2019,HaimPhysRep2019} and experiments involve parallel wires coupled to superconductors~\cite{Baba2018Jun,Ueda2019Oct,Kanne2021Nov,VekrisJJ2021,Vekris2021Sep,Kurtossy2021Oct}. These devices constitute the basis for the design of the so-called Majorana-Cooper pair boxes \cite{Terhal2012PRL,Plugge2016PRB,Plugge2017NJP} and provide the key element for the observation of the predicted topological Kondo effect in hybrid superconductor-double-nanowire islands~\cite{Beri2012Oct,Beri2013,Altland2013May}, which would signal the non-local nature of Majorana zero-energy modes.


Superconducting islands (SIs)~\cite{PhysRevLett.69.1997,PhysRevLett.70.994,JoyezPRL1994,Averin1992Sep,Eiles1993Mar} in \textit{single} nanowires have been extensively studied in the last years. Subjects explored are quasiparticle relaxation, poisoning lifetimes~\cite{higginbotham2015parity,Albrecht2017Mar,MenardPRB2019} and the evolution of SI Coulomb peaks in a magnetic field \cite{Albrecht2016Mar,Sherman2017Mar,Shen2018Nov,SestoftPRM2018,OfarrellPRL2018,vanVeenPRB2018,CarradAdvMat2020,Shen2021,Fleckenstein2018,Saldana2022Jan} to shed light on potential topological properties~\cite{vanHeckPRB2016}. 
Furthermore, devices exploring the Little-Parks effect \cite{Vaitiekenas2020Mar}, interferometry~\cite{Whiticar2020Jun} and reflectometry techniques~\cite{DavydasPRA2019, Sabonis2019Sep,vanVeenPRB2019} yielded additional insight.
The most commonly studied material system has been InAs/Al but alternative superconductors such as NbTiN, Sn and Pb \cite{vanWoerkom2015Jul,Pendharkar2021Apr, Kanne2021Jul, Bjergfelt2021Dec} or different nanowire materials \cite{SestoftPRM2018, VaitiekenasPRL2018} have been explored. Despite this progress and their multiple applications, SIs coupled to multiple nanowires have not yet been demonstrated.

Here we realize multi-terminal hybrid \textit{double}-nanowire SI devices by utilizing\textit{ in-situ} grown InAs semiconductor double nanowires bridged by a small epitaxially-grown Al superconductor. The devices display different charging energy regimes, which we explore in various two-terminal combinations. In a first device, we measure the magnetic field and temperature dependence of the SI Coulomb peaks, and find a similar dependence in every two-terminal combination. Using a thermal model which describes the free energy difference of even and odd states~\cite{higginbotham2015parity}, we extract consistently similar bound state energies for different two-terminal combinations in two different gate configurations. Our results indicate the presence of a common bound state coupled to the four ends of the nanowires and thus extended across the hybrid SI. In a second device, we observe two-electron charging free of quasiparticle poisoning. 

\begin{figure}[h!]
    \centering
    \includegraphics[width=1\linewidth]{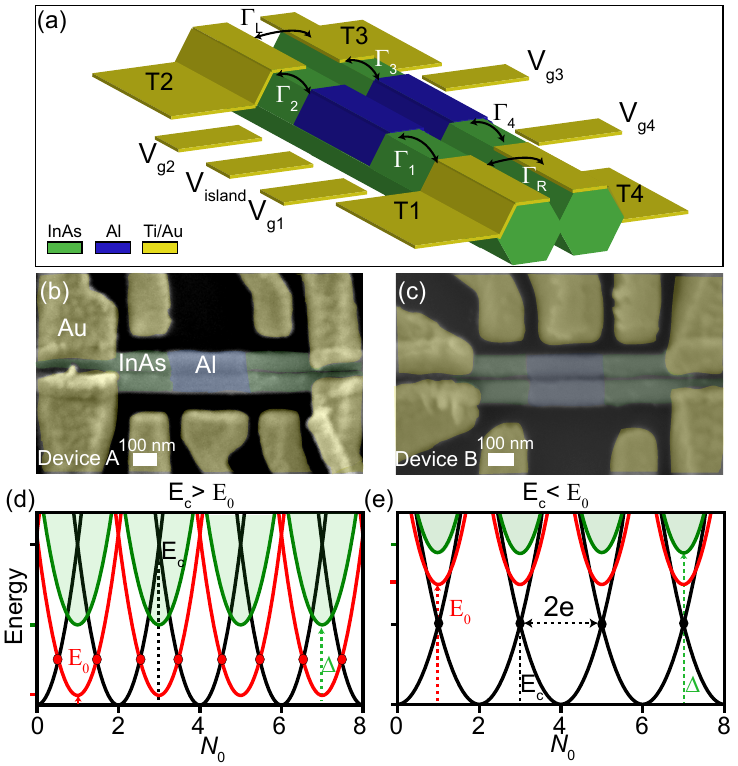}
    \caption{(a) Schematic of a double-nanowire SI device. (b-c) Scanning electron microscopy images of two functional SI devices. (d-e) Energy as a function of the induced charge $N_0 \propto V_{\rm island}$ in the SI for the regime (d) $E_\mathrm{c} > E_0$ and (e) $E_\mathrm{c} < E_0$. Black parabolas describe states with even electron number corresponding to Cooper pairs. Red parabolas describe states with an odd electron number corresponding to a quasiparticle occupying a sub-gap state whose energy at odd $N_0$ values is $E_0$. For $E_\mathrm{c} > E_0$, even-parity states are the ground state in a window of width $2(E_\mathrm{c} + E_0)/E_\mathrm{c}$ centered around even $N_0$ values, while odd-parity states are the ground state in a window of width $2(E_\mathrm{c} - E_0)/E_\mathrm{c}$ centered around odd $N_0$ values. Red dots indicate the crossing of parabolas differing by one electron. For $E_\mathrm{c} < E_0$, even-parity states are the ground state at any $N_0$. The point where two parabolas of the same parity meet yields $E_\mathrm{c}$ measured from the base of the parabolas (black dashed line). Green parabolas correspond to the charge dispersion of the edge of the superconducting gap $\Delta$.}
    \label{fig1}
\end{figure}

\section{Results and discussion}

Figure~\ref{fig1}(a) shows the device concept. Two nanowires (green) provide access to the SI (blue) through four metallic terminals (T1$-$T4, yellow) with coupling tunability ($\Gamma_\mathrm{1-4}$) and gate-induced ($V_{\rm island}$) island charge tunability. Interwire couplings ($\Gamma_\mathrm{L,R}$) are also present. Figure~\ref{fig1}(b,c) shows scanning electron microscopy (SEM) images of the investigated devices. The nanowires are grown by molecular beam epitaxy on a pre-patterned substrate where the inter-wire spacing and their diameters are well controlled~\cite{Kanne2021Nov}. Their hexagonal cross-sections have a diameter $d\approx 85~\mathrm{nm}$, with three of their facets covered by an epitaxially-grown aluminum film of thickness $\approx $17~nm which also bridges the two nanowires. The nanowires are placed on a doped $\mathrm{Si/SiO_2}$ substrate with an oxide thickness of 275~nm and standard electron beam lithography (EBL) techniques are followed to selectively etch the aluminum (using Transene D for 9 seconds) and form 300~nm-long SIs. Contacts and gates are defined by EBL, and metal evaporation of 5~nm~Ti/205~nm~Au is performed after an argon milling treatment of the native InAs oxide to establish ohmic contact. The contacts are separated from the SI by $\approx220$ nm of bare nanowire segments.

The SI is characterized by a capacitance $C$, resulting in a sizable charging energy $E_\mathrm{c} =\frac{e^2}{2C}$. This yields a parabolic energy dispersion of all states against the gate-induced charge $N_0= C_0 V_{\rm island}$ on the island, where $C_0$ is the capacitance between the island and the gate with applied voltage $V_{\rm island}$~\cite{higginbotham2015parity,Albrecht2016Mar} [see Fig. \ref{fig1}(d-e)]. These states are characterized by even or odd occupation numbers. Even states (black lines) correspond to Cooper pair states without Bogoliubov quasiparticle excitations. Odd states at low energies, instead, correspond to the presence of a single Bogoliubov excitation. In particular, for energies above the superconducting gap $\Delta$, the system presents a continuum of one-quasiparticle states (green parabolas). In a SI made of a single conducting material, the ratio $E_\mathrm{c}/\Delta$ determines whether, upon sweeping $V_{\rm island}$, the SI can be filled with electrons one-by-one ($E_\mathrm{c}>\Delta$)~\cite{PhysRevLett.70.994}, or in steps of a Cooper pair ($E_\mathrm{c}<\Delta$)~\cite{PhysRevLett.69.1997}. In our heterostructured devices, the proximity between the semiconducting nanowire (InAs) and the superconducting aluminum gives rise to additional hybridized states. If any of them lie at an energy $E_0$ below $\Delta$ (red parabolas), the charging mechanism is determined by the ratio $E_\mathrm{c}/E_0$ instead. The corresponding parabolic dispersion for the two limiting ratios of $E_\mathrm{c}/E_0$ is depicted in Fig.~\ref{fig1}(d,e) as a function of $N_0$.

A measurable electrical current through the SI is caused by single quasiparticles that excite the ground state into a higher-energy state with charge differing by one electron. Therefore, a zero-bias differential conductance arises when parabolas of different color cross. If Andreev reflections are allowed by sufficient coupling of the SI to the metallic leads, a current is also observed when two black parabolas cross in the $E_\mathrm{c}<E_0$ case in Fig.~\ref{fig1}(e).

\subsection{Zero-bias conductance}

\begin{figure}[th!]
    \centering
    \includegraphics[width=1\linewidth]{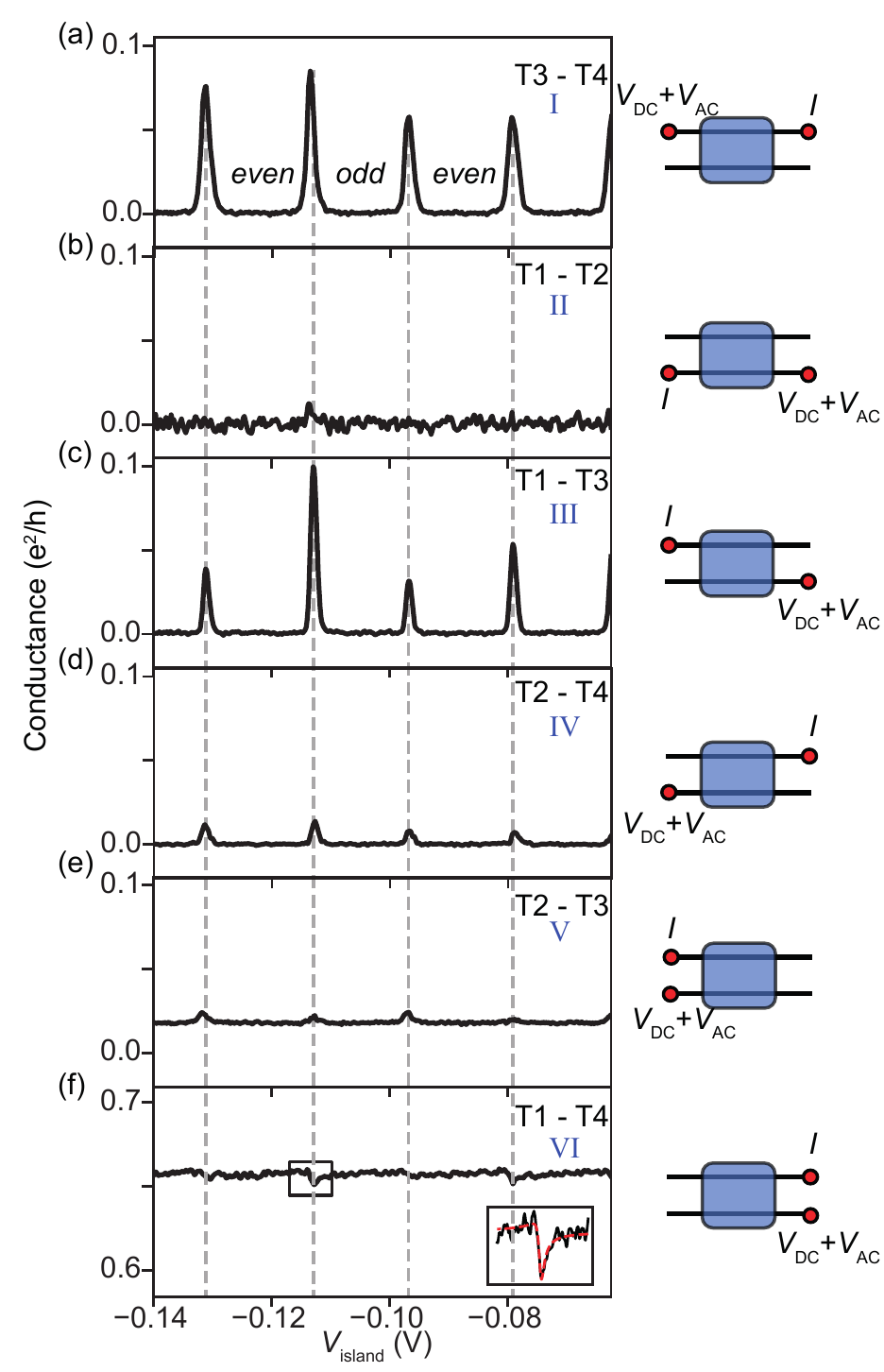}
    \caption{(a-f) Zero-bias conductance traces as a function of $V_\mathrm{island}$ for each two-terminal combination on device A. Each panel is accompanied with a device schematic illustrating the measurement setup. Setups (a,c-f) show clear modulations of the conductance monitored at identical gate voltages (gray dashed lines), proving that the same charged object is probed. Dips of conductance superimposed on a large background conductance in (f) are interpreted as Fano resonances due to the interference of a continuum of states (interwire tunnelling) with the levels of the SI when they are on resonance. The inset shows a fit to the Fano lineshape.}
    \label{fig2}
\end{figure}

In double-nanowire devices with $E_{\rm c} > E_0$, the presence of a sub-gap state sufficiently tunnel-coupled to all leads causes a characteristic transport signature: The zero-bias differential conductance ($\mathrm{dI/dV}$) between any given pair of the four terminals of the device must indeed reproduce the same spacing as a function of $V_{\rm island}$, corresponding to electron loading/unloading into this state at the red dot degeneracies in Fig.~\ref{fig1}(d). To this end, we measure $\mathrm{dI/dV}$ in device A in Fig.~\ref{fig1}(b). It exhibits even-odd charging behavior of the SI consistent with $E_\mathrm{c} > E_0$, in six different two-terminal setups ($\mathrm{I-VI}$) against $V_\mathrm{island}$, as shown in Fig.~\ref{fig2}. The measurements are conducted by sourcing one terminal with an AC voltage of $V_\mathrm{AC}=$~5~$\mu\mathrm{V}$ superimposed on a DC voltage $V_\mathrm{DC}$, and recording the differential conductance $\mathrm{dI/dV}$ on a second terminal while electrically floating the remaining two terminals.
For example, Fig.~\ref{fig2}(a) shows a zero-bias measurement using setup $\mathrm{I}$, where the SI is probed via the upper nanowire through leads T3 and T4, while leads T1 and T2 are floating. Setups $\mathrm{I,III,IV}$ and $\mathrm{V}$ similarly show clear peaks of conductance, while setups $\mathrm{II,VI}$ do not. The visible peaks appear at the same $V_\mathrm{island}$ voltages, which would be the case for shorted end contacts (setup V and VI). However, the lack of additional peaks in the conductance spectra gives a first indication that the Coulomb resonances are consistent with a common sub-gap state in the SI as represented in the simple model of Fig.~\ref{fig1}d.

We ascribe the lack of Coulomb peaks in setup $\mathrm{II}$ to the large coupling asymmetry of terminals T1,T2, which results in very faint features [for high-bias measurements see Supplemental Material (SM)]. With setup $\mathrm{VI}$, a large background conductance of more than 0.6~$\mathrm{e^2/h}$ is measured, reflecting the large direct tunneling $\Gamma_{R}$ between leads T1,T4. A closer inspection reveals asymmetric dips of conductance at the gate values where we observe peaks in the other setups. The dips may be owed to the Fano effect \cite{Fano}, coming from a resonant level from the SI and a highly transparent path with a continuum of states (tunnelling from lead T1 to T4). An inset in Fig.~\ref{fig2}(f) shows a conductance dip fitted with the Fano lineshape yielding $q=-0.55$ consistent with the expected value~\cite{Fano,Fanosinglewall}. Setup $\mathrm{V}$ exhibits a weaker background conductance (0.02~$\mathrm{e^2/h}$), indicating that $\Gamma_{L}<\Gamma_{R}$. Note that all combinations involving lead T2 exhibit weakly conducting peaks suggesting a high asymmetry between the coupling $\Gamma_\mathrm{2}$ and the other couplings $\Gamma_\mathrm{1,3,4}$.

\begin{figure}[h!]
    \centering
    \includegraphics[width=1\linewidth]{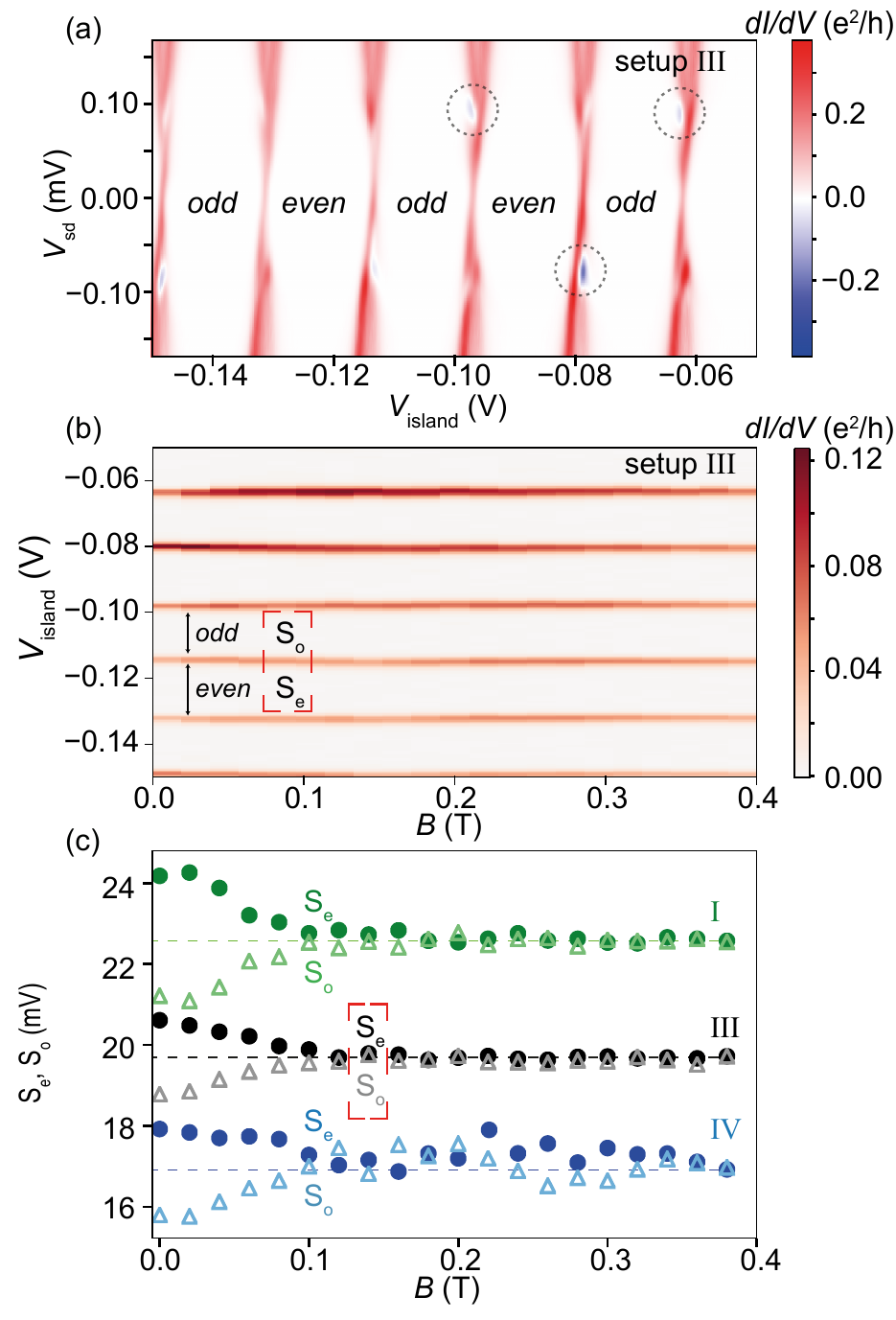}
    \caption{(a) High resolution bias spectroscopy of the SI in setup $\mathrm{III}$ measured at base temperature $T=\mathrm{30~mK}$. Negative differential conductance features (pinpointed by dashed circles) are observed when there is an unpaired quasiparticle in the SI (odd sectors). (b) Magnetic field $B$ dependence of the island resonances measured at zero-bias. Spacings of even and odd sectors are indicated with $\mathrm{S_e}$,$\mathrm{S_o}$. (c) $\mathrm{S_e}$ (circles) and $\mathrm{S_o}$ (hollow triangles) as a function of $B$ for three different measurement setups. Red brackets correspond to the $\mathrm{S_e}$, $\mathrm{S_o}$ spacings marked in (b). Error bars are smaller than the data points, therefore not shown.  Data points of setups $\mathrm{I}$ and $\mathrm{III}$ have been offset on the y-axis for clarity and dashed lines are added as guide to the eye. The low peak conductance in setup $\mathrm{IV}$ makes the peak spacing $\mathrm{S_e}$, $\mathrm{S_o}$ more susceptible to noise. }
    \label{fig3}
\end{figure}

\begin{figure}[h!]
    \centering
   \includegraphics[width=1\linewidth]{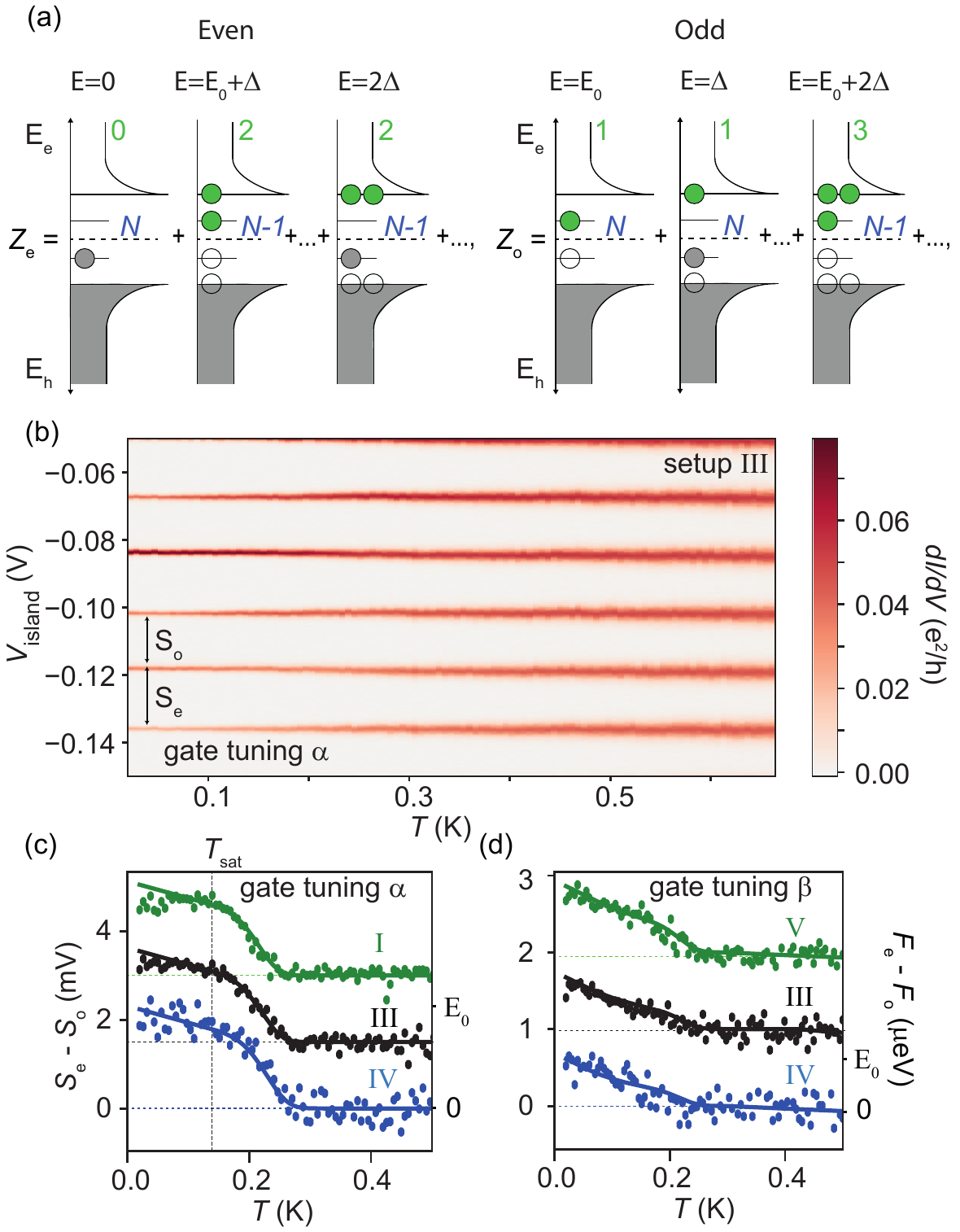}
    \caption{(a) Partition function elements $Z_\mathrm{e}$ and $Z_\mathrm{o}$ for even and odd occupation of the island, respectively. The first term in the left (right) schematic sum corresponds to the case of zero (one) quasiparticle (QP) on the sub-gap state and $N$ number of Cooper pairs on the island. The electron(hole)-like excitation of a QP is pictured as a green solid circle (hollow circle). QP excitations at the gap edge ($\pm \Delta$) or to the sub-gap state $E_0$ are related to breaking a Cooper pair ($N \rightarrow N-1$). Green numbers refer to the total number of QP for a given configuration. Blue labels, $N$ and $N-1$, correspond to the number of Cooper pairs. Energies on each schematic correspond to the energy cost required to realize the respective configuration. Gray color represents filled states. Non-degenerate sub-gap state is assumed. (b) Temperature dependence of island resonances measured in setup $\mathrm{III}$ and gate tuning $\alpha$.  Double headed arrows indicate the size of the analyzed even and odd sectors. (c,d) Difference of even-odd sectors as a function of temperature for three different setups measured in gate tuning $\alpha$ and $\beta$ respectively. Solid lines are fits of the free energy difference using Eq.~S1.}

    \label{fig4}
\end{figure}

\subsection{Bias spectroscopy and magnetic field dependence}

 We estimate $E_\mathrm{c}\approx 0.5~\mathrm{meV}$, (see SM Fig.~S1) and $\Delta=0.19~\mathrm{meV}$ (measured in nanowires from the same growth~\cite{VekrisJJ2021}), from where $E_\mathrm{c} > \Delta$ is confirmed. Note that the $E_\mathrm{c} > \Delta$ case without a sub-gap state $E_0$ would also be consistent with 
 the same Coulomb peaks structure of Fig.~\ref{fig2}. To discern between these two interpretations we perform further measurements. We first focus on a broader bias voltage range. Figure~\ref{fig3}(a) shows a colormap of $\mathrm{dI/dV}$ against bias voltage $V_\mathrm{sd}$ and $V_\mathrm{island}$ measured with setup $\mathrm{III}$. Here, apart from the evident Coulomb blockade structure of the SI, negative differential conductance (NDC) features are observed in blue at finite bias in an alternating pattern. While we can not clearly distinguish all excitation lines, these features resemble previous studies in SIs~\cite{higginbotham2015parity,Albrecht2017Mar}, and appear only at gate ranges where the SI is charged with an $odd$ number of electrons. The NDC appears due to a quasiparticle being trapped in a state weakly coupled to the leads, thus blocking transport through the system~\cite{higginbotham2015parity}. The NDC features appear at a voltage bias smaller than what we expect if they were associated to a quasiparticle trapped above the SC gap, suggesting that more than one subgap states might be present in the device.

Next, we investigate the magnetic field ($B$) dependence of the zero-bias Coulomb peaks in the same $V_\mathrm{island}$ range as Figs.~\ref{fig2} and \ref{fig3}a. In this way, we can establish that the $g$-factor extracted from the Zeeman shift of the Coulomb peaks is consistent with a sub-gap state, and we can confirm the parity assigned to the charge sectors. Figure~\ref{fig3}(b) shows a colormap of the zero-bias differential conductance versus $V_\mathrm{island}$ and $B$ recorded with setup III. Peaks delimiting odd parity sectors split apart with $B$, while those delimiting even parity domains come together. This observation is qualitatively consistent with both $\Delta$ reduction due to pair breaking and Zeeman splitting of the sub-gap state of energy $E_0$ (at $B=0$). To discern between these two effects, we perform a quantitative analysis of the peak spacings from even and odd sectors, $\mathrm{S_e}$ and $\mathrm{S_o}$. The effective $g$-factor of the spacing versus $B$ is estimated at $g=\mathrm{7.8\pm0.2}$, consistent with a sub-gap state induced by the hybridization between the InAs nanowires ($g=-15$ in bulk InAs) and the Al ($g=2$ in bulk Al). To demonstrate that the Zeeman splitting is observed in various setups in consistency with a common sub-gap state in the SI, we show in Fig.~\ref{fig3}(c) the evolution in $B$ of the peak spacing for three different setups measured in the same $V_\mathrm{island}$ range. At $B\approx 100~\mathrm{mT}$, the spacings converge and faint (irregular) oscillations are observed for higher magnetic fields. The convergence of $S_{e,o}$ below the critical field has been analyzed in terms of Andreev states and Majorana zero modes~\cite{Prada2020Oct,Chiu2017,Shen2021,Yu-HuaPRB2021} or the simultaneous diminishing of the gap \cite{Jellinggaard2016Aug}. 
Note that the complete closing of the superconducting gap occurs at higher magnetic field ($B_c\geq 200~\mathrm{mT}$), as determined from additional measurements shown in Fig.~S6 of the SM, which rules out that the extracted effective $g$-factor is due to gap closure only. This lower bound for $B_c$ is in agreement with the upper bound of $B_c=400$ mT obtained from Josephson devices based on identical nanowires~\cite{VekrisJJ2021}.

\begin{table}
    \centering
    \caption{Extracted parameters using eq.~S1 for two different gate tunings of Device A, as shown in Fig.~\ref{fig4}(c,d). The gate voltages for tuning $\alpha$ are $V_\mathrm{g1}=0.35~\mathrm{V}$, $V_\mathrm{g2}=0.5~\mathrm{V}$, $V_\mathrm{g3}=-1~\mathrm{V}$, $V_\mathrm{g4}=2.2~\mathrm{V}$ and $V_\mathrm{bg}=1~\mathrm{V}$. For gate tuning $\beta$ the gates are set at the same voltages except for $V_\mathrm{g3}=-0.8~\mathrm{V}$ and $V_\mathrm{g4}=2.3~\mathrm{V}$. }
    \label{tab1}
    \begin{tabular}{c|ccc}
        \hline
        \multirow{5}{*}{Gate tuning $\alpha$} & \multirow{1}{*}{Setup} & 
         $\Delta(\mu \mathrm{eV})$    & $E_0(\mu \mathrm{eV})$  \\ \hline
       & $\mathrm{I}$             & 171$\pm$3  & 34$\pm$1  \\
       & $\mathrm{III}$             & 175$\pm$3 & 34$\pm$1  \\
       & $\mathrm{IV}$                & 183$\pm$3 & 37$\pm$1  \\ \hline
        \multirow{3}{*}{Gate tuning $\beta$}  
       & $\mathrm{V}$             & 172$\pm$3 & 15$\pm$1   \\
       & $\mathrm{III}$            & 175$\pm$3 & 12$\pm$1 \\
      & $\mathrm{IV}$                 & 173$\pm$5 & 11$\pm$1 \\ \hline  
    \end{tabular}
\end{table}

\subsection{Temperature analysis of the parity effect}

A third piece of evidence of the presence of a common sub-gap state in the SI lies on the temperature evolution of the zero-bias resonances~\cite{higginbotham2015parity}. The temperature modifies the free energy difference between the odd and even states
as $F_{\rm o} - F_{\rm e} = -\mathrm{k_B} T \ln\left(\frac{Z_\mathrm{o} (T)}{Z_\mathrm{e} (T)}\right)$, where $Z_\mathrm{o} (T)$, $Z_\mathrm{e} (T)$ are the partition functions for odd and even states. In Fig.~\ref{fig4}(a), we show the first terms of each partition function, corresponding to the ground state and the first excited states for even and odd occupation of the SI. Excited configurations occur by breaking a Cooper pair and exciting the electrons at the continuum edge or at the sub-gap state ($N\rightarrow N-1$, where $N$ is the number of Cooper pairs), or by exciting quasiparticles from $E_0$ to the continuum edge. Naturally, configurations that involve a large number of excited electrons cost more energy. At low temperatures, only the first terms of the partition function play a role. Thus, $Z_\mathrm{e} > Z_\mathrm{o}$ yielding a finite $F_\mathrm{o} - F_\mathrm{e}$ value which remains nearly constant (saturated) up to a temperature $T_\mathrm{sat}$. As the temperature is increased, the thermal excitation of quasiparticles to the continuum above the superconducting gap is gradually favored, reducing the relative weight of the first terms of $Z_\mathrm{e}$ and $Z_\mathrm{o}$ and increasing the relative weight of higher-order terms which, above $T_\mathrm{sat}$, leads to a linear reduction $F_o - F_e$. Eventually, $Z_\mathrm{e}=Z_\mathrm{o}$, such that $F_\mathrm{o} = F_\mathrm{e}$.

We observe this two-sloped temperature dependence in all two-terminal setups in which Coulomb peaks are visible. Figure \ref{fig4}(b) shows an example colormap of zero-bias differential conductance versus temperature and $V_\mathrm{island}$, the latter swept in the same range as Figs.~\ref{fig2} and \ref{fig3}. The Coulomb peak widths, which are related to all four couplings and the Fermi distribution of the metallic leads, increase with temperature (full-width at half-maximum: 70$~\mu \mathrm{eV}$ at $T$=30~mK; 290$~\mu \mathrm{eV}$ at $T$=0.67~K) as expected. In order to assess the temperature evolution of $F_\mathrm{o} - F_\mathrm{e}$ in this device, we extract the difference in peak spacings $S_\mathrm{e}-S_\mathrm{o}$ versus temperature from this colormap and from similar measurements with two-terminal setups I and IV. The results are plotted in Fig.~\ref{fig4}(c) with vertical offsets for clarity. The two-sloped temperature evolution of $S_\mathrm{e}-S_\mathrm{o}$ culminating in $S_\mathrm{e}=S_\mathrm{o}$ is evident; the initial spacing difference is $\approx2~\mathrm{mV}$ (corresponding to a free energy difference $\Delta F\approx 35~\mu eV$) and the spacings become indistinguishable ($S_\mathrm{e}=S_\mathrm{o}$) at $T\approx 280~\mathrm{mK}$. We fit the data (solid lines) using $S_\mathrm{e}-S_\mathrm{o}=\frac{4}{\alpha e}(F_\mathrm{o}-F_\mathrm{e})$, where $\alpha$ is the lever arm and the difference of the free energies is given by Eq.~S1 in the SM. From the fit, we extract an estimate of $\Delta$ and $E_0$ of the same order for all three setups (see Table~\ref{tab1} for details). This indicates that the same sub-gap state is probed by the different two-terminal combinations. The fit is good for peak spacing data measured at different magnetic fields, which change the value of $E_0$ (see Fig.~S6 in SM). By slightly changing the gate configuration of the device (gate tuning $\beta$), the two slopes in the temperature dependence of $S_\mathrm{e}-S_\mathrm{o}$ become similar (Fig.~\ref{fig4}(d)). The fit to data taken with three different setups yields a lower sub-gap state energy $E_0=13\pm3~\mu \mathrm{eV}$, from which we conclude that $E_0$ is also gate-tunable.

Since the onset ($T_\mathrm{sat}$) of the fast reduction of ($S_\mathrm{e}-S_\mathrm{o}$) with temperature indicates  quasiparticle population above the superconducting gap, previous studies have linked it with the number of quasiparticles $n_\mathrm{qp}V_\mathrm{Al}$ in the SI~\cite{higginbotham2015parity}, where $n_\mathrm{qp}$ is the quasiparticle density and $V_\mathrm{Al}$ the volume of the aluminum island. In our device, $T_\mathrm{sat} \approx 140$ mK, from which we extract $n_\mathrm{qp}V_\mathrm{Al}<3\cdot 10^{-3}$ (see SM for details), in agreement with previous findings for single-nanowire SIs~\cite{higginbotham2015parity}.

 \begin{figure}[h!]
    \centering
    \includegraphics[width=1\linewidth]{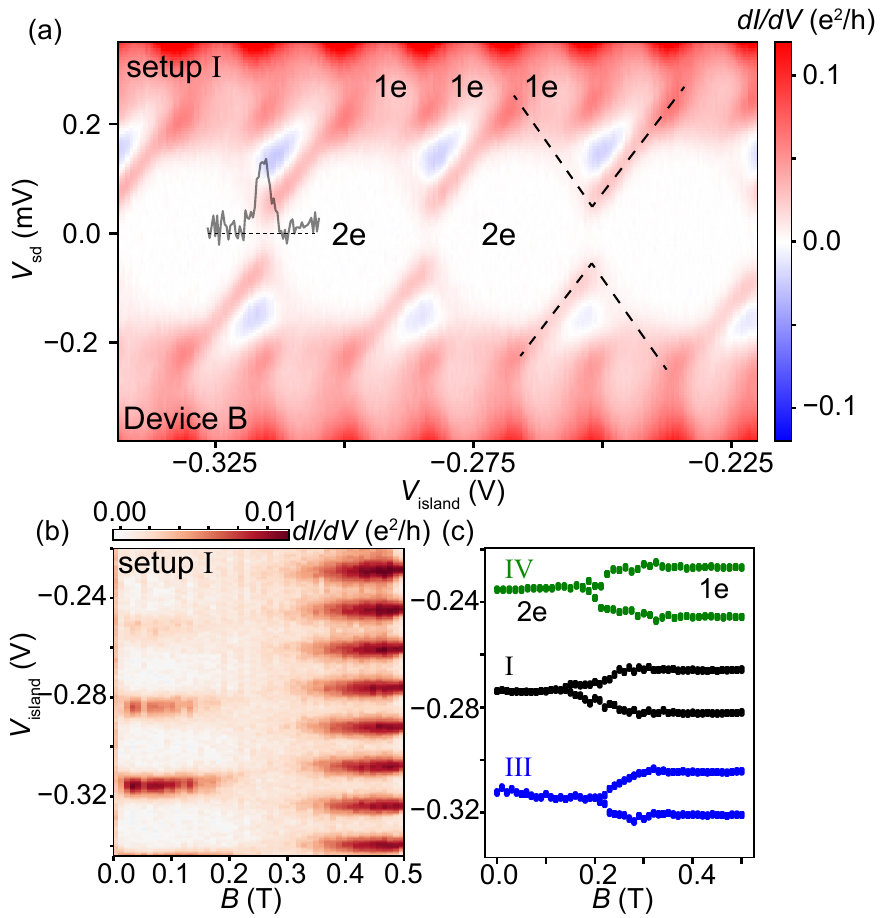}
    \caption{(a) Bias spectroscopy of device B in the $E_\mathrm{c}< E_0$ regime. Two-electron periodicity is observed at low bias while one-electron periodicity is recovered at large bias. The overlaid black trace corresponds to the zero-bias conductance on the given gate segment. The peak conductance is 0.007~$\mathrm{e^2/h}$. (b) Zero-bias magnetic field dependence of the island resonances. Two-electron periodicity is transformed to one-electron periodicity due to the island becoming normal. (c) Peak positions for three different two-terminal setups showing the two-electron to one-electron transition of the SI. }
    \label{fig5}
\end{figure}

\subsection{Two-electron charging in device B}

Finally, we report measurements of the additional device shown in Fig.~\ref{fig1}(c) where a smaller charging energy $E_\mathrm{c}$ of the SI (caused by larger capacitances to the gates and electrodes) allows for Cooper-pair (2e) charging, rather than 1e periodicity shown above. Figure~\ref{fig5}(a) presents a bias spectroscopy measurement using setup $\mathrm{I}$, from which $E_\mathrm{c}\approx 55~\mu \mathrm{eV}$ and $E_0 \approx 65~\mu \mathrm{eV}$ are extracted, and therefore $E_\mathrm{c}<E_0$. Single-electron excitations between even and odd charge parabolas (Fig.~\ref{fig1}e) result in Coulomb diamond features where the lower tips of the diamonds do not touch zero bias (see dashed lines). However, due to significant couplings to the normal leads, a zero-bias signal is visible, as shown by the black trace taken along one of the crossings. The spacing between these faint zero-bias peaks is $2e$ periodic, indicating Cooper-pair charging in this regime. To confirm the $2e$ periodicity of the zero-bias peaks, we investigate their $B$ dependence. An example of a two-terminal measurement using setup~I is shown in Fig.~\ref{fig5}(b), while Fig.~\ref{fig5}(c) shows extracted peak positions versus $B$ for three different two-terminal measurement setups. The $2e$-periodic resonances split and turn into $1e$-periodic ones at $B\approx 320~\mathrm{mT}$ for all setups, at which point $B \approx B_c$. No indications of quasiparticle poisoning are observed at low $B$~\cite{Albrecht2017Mar}. It is worth noting that the interwire tunneling $\Gamma_{\mathrm{L,R}}$ in Device B is lower than Device A (see SM, Fig.~S10), consistent with the larger spacing between the nanowires, as seen in Fig.~\ref{fig1}(c).

\section{Conclusions}

In this work we have characterized SI devices fabricated in an in-situ aluminum deposited double-nanowire platform. The main virtue of our double-nanowire SI devices is the presence of multiple terminals. We performed two-terminal transport measurements, similar to those used for single-nanowire SI devices, using combinations of the multiple terminals of device A. The various two-terminal measurement combinations indicate the presence of a sub-gap state extended across the hybrid DNW SI as follows. First, Coulomb peaks with even-odd ground state parity alternation were observed at the same gate voltages in all setups where they were visible, indicating coupling to a common sub-gap state, however, with a background leakage between the end-end contacts preventing fully independent two-terminal setups.
Second, NDC was observed in bias spectroscopy indicating the blocking of a quasiparticle relaxed by a sub-gap state. Third, the $g$-factor of the shifts of Coulomb peaks with $B$ was estimated to be larger than that of pure Al, consistent with the shifts being due to the Zeeman splitting of a hybridized sub-gap state. Finally, a fit of the temperature dependence of the Coulomb peaks with a thermal model was used to extract the energy of the sub-gap state $E_0$ and the superconducting gap $\Delta$ for various two-terminal measurement setups, finding common values. The fit was done for a second gate setting ($\beta$), obtaining a common $E_0$ value as well, and the same $\Delta$ value as in the first setting ($\alpha$), with the difference in $E_0$ in the two settings attributed to the influence of the gate on the InAs-Al hybridization~\cite{PhysRevX.8.031040}. The presence of independent sub-gap states in each NW would be distinguished by observing different energies in certain two-terminal configurations.

Delocalized states can provide interesting long-range couplings in hybrid devices. For example, a quasiparticle occupying a delocalized state can be used to couple localized spins in quantum dots via the Yu-Shiba-Rusinov mechanism across distances as long as the SI length~\cite{Saldana2022Jan}.

Our devices exhibited different charging regimes, which enabled the study of the quasiparticle population with different methods. In Device A ($E_\mathrm{c}>\Delta$ regime), the quasiparticle density was found to be comparable to that of single-nanowire SI devices~\cite{higginbotham2015parity}. In Device B ($E_\mathrm{c}<\Delta$), no signs of quasiparticle poisoning were observed. Our double-nanowire platform thus shows promise for applications where parity conservation is required~\cite{Albrecht2016Mar}.

Despite the challenges of interwire coupling and low $B_c$, our devices comprise a step towards the realization of exotic proposals in parallel nanowires coupled by superconductors~\cite{Beri2012Oct,Beri2013,Altland2013May,Klinovaja2014Jul,SchradePRB2017} and alternative systems for multichannel Kondo phenomena \cite{IftikharNat2015,IftikharScience2018,pousearxiv2021}. The above limitations can be overcome by employing nanowires well separated from one another integrated by higher $B_c$ superconductors such as Pb~\cite{Kanne2021Jul}. Such advances would further our understanding of the sub-gap states (e.g.\ Yu-Shiba-Rusinov, Andreev bound states or Majorana zero-modes) hosted by parallel nanowires bridged by SIs. In particular, a new generation of devices showing $1e$ charging behavior in a large field range below $B_c$ could shed light on the nature of the involved sub-gap state by investigating the predictions related to multichannel or topological Kondo phenomena.

\section{Acknowledgments}
We thank Karsten Flensberg for valuable discussions. We acknowledge the support of the Sino-Danish Center, Villum Foundation (Research Grant No. 25310), the European Union’s Horizon 2020 research and innovation program under the Marie Sklodowska-Curie Grant Agreements No. 832645 "SpinScreen" and No. 847523 "INTERACTIONS", QuantERA SuperTop (NN 127900), the European Union’s Horizon 2020 research and innovation programme FETOpen Grant No. 828948 (AndQC), the Danish National Research Foundation, Carlsberg Foundation and the Independent Research Fund Denmark.

\bibliography{bibl}

\providecommand{\noopsort}[1]{}\providecommand{\singleletter}[1]{#1}%
\begin{thebibliography}{65}%
\makeatletter
\providecommand \@ifxundefined [1]{%
 \@ifx{#1\undefined}
}%
\providecommand \@ifnum [1]{%
 \ifnum #1\expandafter \@firstoftwo
 \else \expandafter \@secondoftwo
 \fi
}%
\providecommand \@ifx [1]{%
 \ifx #1\expandafter \@firstoftwo
 \else \expandafter \@secondoftwo
 \fi
}%
\providecommand \natexlab [1]{#1}%
\providecommand \enquote  [1]{``#1''}%
\providecommand \bibnamefont  [1]{#1}%
\providecommand \bibfnamefont [1]{#1}%
\providecommand \citenamefont [1]{#1}%
\providecommand \href@noop [0]{\@secondoftwo}%
\providecommand \href [0]{\begingroup \@sanitize@url \@href}%
\providecommand \@href[1]{\@@startlink{#1}\@@href}%
\providecommand \@@href[1]{\endgroup#1\@@endlink}%
\providecommand \@sanitize@url [0]{\catcode `\\12\catcode `\$12\catcode
  `\&12\catcode `\#12\catcode `\^12\catcode `\_12\catcode `\%12\relax}%
\providecommand \@@startlink[1]{}%
\providecommand \@@endlink[0]{}%
\providecommand \url  [0]{\begingroup\@sanitize@url \@url }%
\providecommand \@url [1]{\endgroup\@href {#1}{\urlprefix }}%
\providecommand \urlprefix  [0]{URL }%
\providecommand \Eprint [0]{\href }%
\providecommand \doibase [0]{https://doi.org/}%
\providecommand \selectlanguage [0]{\@gobble}%
\providecommand \bibinfo  [0]{\@secondoftwo}%
\providecommand \bibfield  [0]{\@secondoftwo}%
\providecommand \translation [1]{[#1]}%
\providecommand \BibitemOpen [0]{}%
\providecommand \bibitemStop [0]{}%
\providecommand \bibitemNoStop [0]{.\EOS\space}%
\providecommand \EOS [0]{\spacefactor3000\relax}%
\providecommand \BibitemShut  [1]{\csname bibitem#1\endcsname}%
\let\auto@bib@innerbib\@empty
\bibitem [{\citenamefont {Aasen}\ \emph {et~al.}(2016)\citenamefont {Aasen},
  \citenamefont {Hell}, \citenamefont {Mishmash}, \citenamefont {Higginbotham},
  \citenamefont {Danon}, \citenamefont {Leijnse}, \citenamefont {Jespersen},
  \citenamefont {Folk}, \citenamefont {Marcus}, \citenamefont {Flensberg},\
  and\ \citenamefont {Alicea}}]{Aasen2016Aug}%
  \BibitemOpen
  \bibfield  {author} {\bibinfo {author} {\bibfnamefont {D.}~\bibnamefont
  {Aasen}}, \bibinfo {author} {\bibfnamefont {M.}~\bibnamefont {Hell}},
  \bibinfo {author} {\bibfnamefont {R.~V.}\ \bibnamefont {Mishmash}}, \bibinfo
  {author} {\bibfnamefont {A.}~\bibnamefont {Higginbotham}}, \bibinfo {author}
  {\bibfnamefont {J.}~\bibnamefont {Danon}}, \bibinfo {author} {\bibfnamefont
  {M.}~\bibnamefont {Leijnse}}, \bibinfo {author} {\bibfnamefont {T.~S.}\
  \bibnamefont {Jespersen}}, \bibinfo {author} {\bibfnamefont {J.~A.}\
  \bibnamefont {Folk}}, \bibinfo {author} {\bibfnamefont {C.~M.}\ \bibnamefont
  {Marcus}}, \bibinfo {author} {\bibfnamefont {K.}~\bibnamefont {Flensberg}},\
  and\ \bibinfo {author} {\bibfnamefont {J.}~\bibnamefont {Alicea}},\
  }\bibfield  {title} {\bibinfo {title} {{Milestones Toward Majorana-Based
  Quantum Computing}},\ }\href {https://doi.org/10.1103/PhysRevX.6.031016}
  {\bibfield  {journal} {\bibinfo  {journal} {Phys. Rev. X}\ }\textbf {\bibinfo
  {volume} {6}},\ \bibinfo {pages} {031016} (\bibinfo {year}
  {2016})}\BibitemShut {NoStop}%
\bibitem [{\citenamefont {Yao}\ \emph {et~al.}(2014)\citenamefont {Yao},
  \citenamefont {Moca}, \citenamefont {Weymann}, \citenamefont {Sau},
  \citenamefont {Lukin}, \citenamefont {Demler},\ and\ \citenamefont
  {Zar{\ifmmode\acute{a}\else\'{a}\fi}nd}}]{Yao2014Dec}%
  \BibitemOpen
  \bibfield  {author} {\bibinfo {author} {\bibfnamefont {N.~Y.}\ \bibnamefont
  {Yao}}, \bibinfo {author} {\bibfnamefont {C.~P.}\ \bibnamefont {Moca}},
  \bibinfo {author} {\bibfnamefont {I.}~\bibnamefont {Weymann}}, \bibinfo
  {author} {\bibfnamefont {J.~D.}\ \bibnamefont {Sau}}, \bibinfo {author}
  {\bibfnamefont {M.~D.}\ \bibnamefont {Lukin}}, \bibinfo {author}
  {\bibfnamefont {E.~A.}\ \bibnamefont {Demler}},\ and\ \bibinfo {author}
  {\bibfnamefont {G.}~\bibnamefont {Zar{\ifmmode\acute{a}\else\'{a}\fi}nd}},\
  }\bibfield  {title} {\bibinfo {title} {{Phase diagram and excitations of a
  Shiba molecule}},\ }\href {https://doi.org/10.1103/PhysRevB.90.241108}
  {\bibfield  {journal} {\bibinfo  {journal} {Phys. Rev. B}\ }\textbf {\bibinfo
  {volume} {90}},\ \bibinfo {pages} {241108(R)} (\bibinfo {year}
  {2014})}\BibitemShut {NoStop}%
\bibitem [{\citenamefont
  {K{\ifmmode\ddot{u}\else\"{u}\fi}rt{\ifmmode\ddot{o}\else\"{o}\fi}ssy}\ \emph
  {et~al.}(2021)\citenamefont
  {K{\ifmmode\ddot{u}\else\"{u}\fi}rt{\ifmmode\ddot{o}\else\"{o}\fi}ssy},
  \citenamefont {Scher{\ifmmode\ddot{u}\else\"{u}\fi}bl}, \citenamefont
  {F{\ifmmode\ddot{u}\else\"{u}\fi}l{\ifmmode\ddot{o}\else\"{o}\fi}p},
  \citenamefont {Luk{\ifmmode\acute{a}\else\'{a}\fi}cs}, \citenamefont {Kanne},
  \citenamefont {Nyg{\aa}rd}, \citenamefont {Makk},\ and\ \citenamefont
  {Csonka}}]{Kurtossy2021Oct}%
  \BibitemOpen
  \bibfield  {author} {\bibinfo {author} {\bibfnamefont {O.}~\bibnamefont
  {K{\ifmmode\ddot{u}\else\"{u}\fi}rt{\ifmmode\ddot{o}\else\"{o}\fi}ssy}},
  \bibinfo {author} {\bibfnamefont {Z.}~\bibnamefont
  {Scher{\ifmmode\ddot{u}\else\"{u}\fi}bl}}, \bibinfo {author} {\bibfnamefont
  {G.}~\bibnamefont
  {F{\ifmmode\ddot{u}\else\"{u}\fi}l{\ifmmode\ddot{o}\else\"{o}\fi}p}},
  \bibinfo {author} {\bibfnamefont {I.~E.}\ \bibnamefont
  {Luk{\ifmmode\acute{a}\else\'{a}\fi}cs}}, \bibinfo {author} {\bibfnamefont
  {T.}~\bibnamefont {Kanne}}, \bibinfo {author} {\bibfnamefont
  {J.}~\bibnamefont {Nyg{\aa}rd}}, \bibinfo {author} {\bibfnamefont
  {P.}~\bibnamefont {Makk}},\ and\ \bibinfo {author} {\bibfnamefont
  {S.}~\bibnamefont {Csonka}},\ }\bibfield  {title} {\bibinfo {title} {{Andreev
  Molecule in Parallel InAs Nanowires}},\ }\href
  {https://doi.org/10.1021/acs.nanolett.1c01956} {\bibfield  {journal}
  {\bibinfo  {journal} {Nano Lett.}\ }\textbf {\bibinfo {volume} {21}},\
  \bibinfo {pages} {7929} (\bibinfo {year} {2021})}\BibitemShut {NoStop}%
\bibitem [{\citenamefont {Gaidamauskas}\ \emph {et~al.}(2014)\citenamefont
  {Gaidamauskas}, \citenamefont {Paaske},\ and\ \citenamefont
  {Flensberg}}]{GaidamauskasPRL2014}%
  \BibitemOpen
  \bibfield  {author} {\bibinfo {author} {\bibfnamefont {E.}~\bibnamefont
  {Gaidamauskas}}, \bibinfo {author} {\bibfnamefont {J.}~\bibnamefont
  {Paaske}},\ and\ \bibinfo {author} {\bibfnamefont {K.}~\bibnamefont
  {Flensberg}},\ }\bibfield  {title} {\bibinfo {title} {Majorana bound states
  in two-channel time-reversal-symmetric nanowire systems},\ }\href
  {https://doi.org/10.1103/PhysRevLett.112.126402} {\bibfield  {journal}
  {\bibinfo  {journal} {Phys. Rev. Lett.}\ }\textbf {\bibinfo {volume} {112}},\
  \bibinfo {pages} {126402} (\bibinfo {year} {2014})}\BibitemShut {NoStop}%
\bibitem [{\citenamefont {Klinovaja}\ and\ \citenamefont
  {Loss}(2014)}]{Klinovaja2014Jul}%
  \BibitemOpen
  \bibfield  {author} {\bibinfo {author} {\bibfnamefont {J.}~\bibnamefont
  {Klinovaja}}\ and\ \bibinfo {author} {\bibfnamefont {D.}~\bibnamefont
  {Loss}},\ }\bibfield  {title} {\bibinfo {title} {{Time-reversal invariant
  parafermions in interacting Rashba nanowires}},\ }\href
  {https://doi.org/10.1103/PhysRevB.90.045118} {\bibfield  {journal} {\bibinfo
  {journal} {Phys. Rev. B}\ }\textbf {\bibinfo {volume} {90}},\ \bibinfo
  {pages} {045118} (\bibinfo {year} {2014})}\BibitemShut {NoStop}%
\bibitem [{\citenamefont {Ebisu}\ \emph {et~al.}(2016)\citenamefont {Ebisu},
  \citenamefont {Lu}, \citenamefont {Klinovaja},\ and\ \citenamefont
  {Tanaka}}]{EbisuProg2016}%
  \BibitemOpen
  \bibfield  {author} {\bibinfo {author} {\bibfnamefont {H.}~\bibnamefont
  {Ebisu}}, \bibinfo {author} {\bibfnamefont {B.}~\bibnamefont {Lu}}, \bibinfo
  {author} {\bibfnamefont {J.}~\bibnamefont {Klinovaja}},\ and\ \bibinfo
  {author} {\bibfnamefont {Y.}~\bibnamefont {Tanaka}},\ }\bibfield  {title}
  {\bibinfo {title} {{Theory of time-reversal topological superconductivity in
  double Rashba wires: symmetries of Cooper pairs and Andreev bound states}},\
  }\bibfield  {journal} {\bibinfo  {journal} {Progress of Theoretical and
  Experimental Physics}\ }\textbf {\bibinfo {volume} {2016}},\ \href
  {https://doi.org/10.1093/ptep/ptw094} {10.1093/ptep/ptw094} (\bibinfo {year}
  {2016}),\ \bibinfo {note} {083I01}\BibitemShut {NoStop}%
\bibitem [{\citenamefont {Schrade}\ \emph {et~al.}(2017)\citenamefont
  {Schrade}, \citenamefont {Thakurathi}, \citenamefont {Reeg}, \citenamefont
  {Hoffman}, \citenamefont {Klinovaja},\ and\ \citenamefont
  {Loss}}]{SchradePRB2017}%
  \BibitemOpen
  \bibfield  {author} {\bibinfo {author} {\bibfnamefont {C.}~\bibnamefont
  {Schrade}}, \bibinfo {author} {\bibfnamefont {M.}~\bibnamefont {Thakurathi}},
  \bibinfo {author} {\bibfnamefont {C.}~\bibnamefont {Reeg}}, \bibinfo {author}
  {\bibfnamefont {S.}~\bibnamefont {Hoffman}}, \bibinfo {author} {\bibfnamefont
  {J.}~\bibnamefont {Klinovaja}},\ and\ \bibinfo {author} {\bibfnamefont
  {D.}~\bibnamefont {Loss}},\ }\bibfield  {title} {\bibinfo {title} {Low-field
  topological threshold in majorana double nanowires},\ }\href
  {https://doi.org/10.1103/PhysRevB.96.035306} {\bibfield  {journal} {\bibinfo
  {journal} {Phys. Rev. B}\ }\textbf {\bibinfo {volume} {96}},\ \bibinfo
  {pages} {035306} (\bibinfo {year} {2017})}\BibitemShut {NoStop}%
\bibitem [{\citenamefont {Reeg}\ \emph {et~al.}(2017)\citenamefont {Reeg},
  \citenamefont {Klinovaja},\ and\ \citenamefont {Loss}}]{ReegPRB2017}%
  \BibitemOpen
  \bibfield  {author} {\bibinfo {author} {\bibfnamefont {C.}~\bibnamefont
  {Reeg}}, \bibinfo {author} {\bibfnamefont {J.}~\bibnamefont {Klinovaja}},\
  and\ \bibinfo {author} {\bibfnamefont {D.}~\bibnamefont {Loss}},\ }\bibfield
  {title} {\bibinfo {title} {Destructive interference of direct and crossed
  andreev pairing in a system of two nanowires coupled via an $s$-wave
  superconductor},\ }\href {https://doi.org/10.1103/PhysRevB.96.081301}
  {\bibfield  {journal} {\bibinfo  {journal} {Phys. Rev. B}\ }\textbf {\bibinfo
  {volume} {96}},\ \bibinfo {pages} {081301(R)} (\bibinfo {year}
  {2017})}\BibitemShut {NoStop}%
\bibitem [{\citenamefont {Schrade}\ and\ \citenamefont
  {Fu}(2018)}]{SchradePRL2018}%
  \BibitemOpen
  \bibfield  {author} {\bibinfo {author} {\bibfnamefont {C.}~\bibnamefont
  {Schrade}}\ and\ \bibinfo {author} {\bibfnamefont {L.}~\bibnamefont {Fu}},\
  }\bibfield  {title} {\bibinfo {title} {Parity-controlled $2\ensuremath{\pi}$
  josephson effect mediated by majorana kramers pairs},\ }\href
  {https://doi.org/10.1103/PhysRevLett.120.267002} {\bibfield  {journal}
  {\bibinfo  {journal} {Phys. Rev. Lett.}\ }\textbf {\bibinfo {volume} {120}},\
  \bibinfo {pages} {267002} (\bibinfo {year} {2018})}\BibitemShut {NoStop}%
\bibitem [{\citenamefont {Thakurathi}\ \emph {et~al.}(2018)\citenamefont
  {Thakurathi}, \citenamefont {Simon}, \citenamefont {Mandal}, \citenamefont
  {Klinovaja},\ and\ \citenamefont {Loss}}]{ThakurathiPRB2018}%
  \BibitemOpen
  \bibfield  {author} {\bibinfo {author} {\bibfnamefont {M.}~\bibnamefont
  {Thakurathi}}, \bibinfo {author} {\bibfnamefont {P.}~\bibnamefont {Simon}},
  \bibinfo {author} {\bibfnamefont {I.}~\bibnamefont {Mandal}}, \bibinfo
  {author} {\bibfnamefont {J.}~\bibnamefont {Klinovaja}},\ and\ \bibinfo
  {author} {\bibfnamefont {D.}~\bibnamefont {Loss}},\ }\bibfield  {title}
  {\bibinfo {title} {Majorana kramers pairs in rashba double nanowires with
  interactions and disorder},\ }\href
  {https://doi.org/10.1103/PhysRevB.97.045415} {\bibfield  {journal} {\bibinfo
  {journal} {Phys. Rev. B}\ }\textbf {\bibinfo {volume} {97}},\ \bibinfo
  {pages} {045415} (\bibinfo {year} {2018})}\BibitemShut {NoStop}%
\bibitem [{\citenamefont {Dmytruk}\ \emph {et~al.}(2019)\citenamefont
  {Dmytruk}, \citenamefont {Thakurathi}, \citenamefont {Loss},\ and\
  \citenamefont {Klinovaja}}]{DmytrukPRB2019}%
  \BibitemOpen
  \bibfield  {author} {\bibinfo {author} {\bibfnamefont {O.}~\bibnamefont
  {Dmytruk}}, \bibinfo {author} {\bibfnamefont {M.}~\bibnamefont {Thakurathi}},
  \bibinfo {author} {\bibfnamefont {D.}~\bibnamefont {Loss}},\ and\ \bibinfo
  {author} {\bibfnamefont {J.}~\bibnamefont {Klinovaja}},\ }\bibfield  {title}
  {\bibinfo {title} {Majorana bound states in double nanowires with reduced
  zeeman thresholds due to supercurrents},\ }\href
  {https://doi.org/10.1103/PhysRevB.99.245416} {\bibfield  {journal} {\bibinfo
  {journal} {Phys. Rev. B}\ }\textbf {\bibinfo {volume} {99}},\ \bibinfo
  {pages} {245416} (\bibinfo {year} {2019})}\BibitemShut {NoStop}%
\bibitem [{\citenamefont {Thakurathi}\ \emph {et~al.}(2020)\citenamefont
  {Thakurathi}, \citenamefont {Chevallier}, \citenamefont {Loss},\ and\
  \citenamefont {Klinovaja}}]{ThakurathiPRR2020}%
  \BibitemOpen
  \bibfield  {author} {\bibinfo {author} {\bibfnamefont {M.}~\bibnamefont
  {Thakurathi}}, \bibinfo {author} {\bibfnamefont {D.}~\bibnamefont
  {Chevallier}}, \bibinfo {author} {\bibfnamefont {D.}~\bibnamefont {Loss}},\
  and\ \bibinfo {author} {\bibfnamefont {J.}~\bibnamefont {Klinovaja}},\
  }\bibfield  {title} {\bibinfo {title} {Transport signatures of bulk
  topological phases in double rashba nanowires probed by spin-polarized stm},\
  }\href {https://doi.org/10.1103/PhysRevResearch.2.023197} {\bibfield
  {journal} {\bibinfo  {journal} {Phys. Rev. Research}\ }\textbf {\bibinfo
  {volume} {2}},\ \bibinfo {pages} {023197} (\bibinfo {year}
  {2020})}\BibitemShut {NoStop}%
\bibitem [{\citenamefont {Kotetes}\ \emph {et~al.}(2019)\citenamefont
  {Kotetes}, \citenamefont {Mercaldo},\ and\ \citenamefont
  {Cuoco}}]{KotetesPRL2019}%
  \BibitemOpen
  \bibfield  {author} {\bibinfo {author} {\bibfnamefont {P.}~\bibnamefont
  {Kotetes}}, \bibinfo {author} {\bibfnamefont {M.~T.}\ \bibnamefont
  {Mercaldo}},\ and\ \bibinfo {author} {\bibfnamefont {M.}~\bibnamefont
  {Cuoco}},\ }\bibfield  {title} {\bibinfo {title} {Synthetic weyl points and
  chiral anomaly in majorana devices with nonstandard andreev-bound-state
  spectra},\ }\href {https://doi.org/10.1103/PhysRevLett.123.126802} {\bibfield
   {journal} {\bibinfo  {journal} {Phys. Rev. Lett.}\ }\textbf {\bibinfo
  {volume} {123}},\ \bibinfo {pages} {126802} (\bibinfo {year}
  {2019})}\BibitemShut {NoStop}%
\bibitem [{\citenamefont {Papaj}\ \emph {et~al.}(2019)\citenamefont {Papaj},
  \citenamefont {Zhu},\ and\ \citenamefont {Fu}}]{PapajPRB2019}%
  \BibitemOpen
  \bibfield  {author} {\bibinfo {author} {\bibfnamefont {M.}~\bibnamefont
  {Papaj}}, \bibinfo {author} {\bibfnamefont {Z.}~\bibnamefont {Zhu}},\ and\
  \bibinfo {author} {\bibfnamefont {L.}~\bibnamefont {Fu}},\ }\bibfield
  {title} {\bibinfo {title} {Multichannel charge kondo effect and
  non-fermi-liquid fixed points in conventional and topological superconductor
  islands},\ }\href {https://doi.org/10.1103/PhysRevB.99.014512} {\bibfield
  {journal} {\bibinfo  {journal} {Phys. Rev. B}\ }\textbf {\bibinfo {volume}
  {99}},\ \bibinfo {pages} {014512} (\bibinfo {year} {2019})}\BibitemShut
  {NoStop}%
\bibitem [{\citenamefont {Haim}\ and\ \citenamefont
  {Oreg}(2019)}]{HaimPhysRep2019}%
  \BibitemOpen
  \bibfield  {author} {\bibinfo {author} {\bibfnamefont {A.}~\bibnamefont
  {Haim}}\ and\ \bibinfo {author} {\bibfnamefont {Y.}~\bibnamefont {Oreg}},\
  }\bibfield  {title} {\bibinfo {title} {Time-reversal-invariant topological
  superconductivity in one and two dimensions},\ }\href
  {https://doi.org/https://doi.org/10.1016/j.physrep.2019.08.002} {\bibfield
  {journal} {\bibinfo  {journal} {Physics Reports}\ }\textbf {\bibinfo {volume}
  {825}},\ \bibinfo {pages} {1} (\bibinfo {year} {2019})}\BibitemShut {NoStop}%
\bibitem [{\citenamefont {Baba}\ \emph {et~al.}(2018)\citenamefont {Baba},
  \citenamefont {J{\ifmmode\ddot{u}\else\"{u}\fi}nger}, \citenamefont {Matsuo},
  \citenamefont {Baumgartner}, \citenamefont {Sato}, \citenamefont {Kamata},
  \citenamefont {Li}, \citenamefont {Jeppesen}, \citenamefont {Samuelson},
  \citenamefont {Xu}, \citenamefont
  {Sch{\ifmmode\ddot{o}\else\"{o}\fi}nenberger},\ and\ \citenamefont
  {Tarucha}}]{Baba2018Jun}%
  \BibitemOpen
  \bibfield  {author} {\bibinfo {author} {\bibfnamefont {S.}~\bibnamefont
  {Baba}}, \bibinfo {author} {\bibfnamefont {C.}~\bibnamefont
  {J{\ifmmode\ddot{u}\else\"{u}\fi}nger}}, \bibinfo {author} {\bibfnamefont
  {S.}~\bibnamefont {Matsuo}}, \bibinfo {author} {\bibfnamefont
  {A.}~\bibnamefont {Baumgartner}}, \bibinfo {author} {\bibfnamefont
  {Y.}~\bibnamefont {Sato}}, \bibinfo {author} {\bibfnamefont {H.}~\bibnamefont
  {Kamata}}, \bibinfo {author} {\bibfnamefont {K.}~\bibnamefont {Li}}, \bibinfo
  {author} {\bibfnamefont {S.}~\bibnamefont {Jeppesen}}, \bibinfo {author}
  {\bibfnamefont {L.}~\bibnamefont {Samuelson}}, \bibinfo {author}
  {\bibfnamefont {H.~Q.}\ \bibnamefont {Xu}}, \bibinfo {author} {\bibfnamefont
  {C.}~\bibnamefont {Sch{\ifmmode\ddot{o}\else\"{o}\fi}nenberger}},\ and\
  \bibinfo {author} {\bibfnamefont {S.}~\bibnamefont {Tarucha}},\ }\bibfield
  {title} {\bibinfo {title} {{Cooper-pair splitting in two parallel InAs
  nanowires}},\ }\href {https://doi.org/10.1088/1367-2630/aac74e} {\bibfield
  {journal} {\bibinfo  {journal} {New J. Phys.}\ }\textbf {\bibinfo {volume}
  {20}},\ \bibinfo {pages} {063021} (\bibinfo {year} {2018})}\BibitemShut
  {NoStop}%
\bibitem [{\citenamefont {Ueda}\ \emph {et~al.}(2019)\citenamefont {Ueda},
  \citenamefont {Matsuo}, \citenamefont {Kamata}, \citenamefont {Baba},
  \citenamefont {Sato}, \citenamefont {Takeshige}, \citenamefont {Li},
  \citenamefont {Jeppesen}, \citenamefont {Samuelson}, \citenamefont {Xu},\
  and\ \citenamefont {Tarucha}}]{Ueda2019Oct}%
  \BibitemOpen
  \bibfield  {author} {\bibinfo {author} {\bibfnamefont {K.}~\bibnamefont
  {Ueda}}, \bibinfo {author} {\bibfnamefont {S.}~\bibnamefont {Matsuo}},
  \bibinfo {author} {\bibfnamefont {H.}~\bibnamefont {Kamata}}, \bibinfo
  {author} {\bibfnamefont {S.}~\bibnamefont {Baba}}, \bibinfo {author}
  {\bibfnamefont {Y.}~\bibnamefont {Sato}}, \bibinfo {author} {\bibfnamefont
  {Y.}~\bibnamefont {Takeshige}}, \bibinfo {author} {\bibfnamefont
  {K.}~\bibnamefont {Li}}, \bibinfo {author} {\bibfnamefont {S.}~\bibnamefont
  {Jeppesen}}, \bibinfo {author} {\bibfnamefont {L.}~\bibnamefont {Samuelson}},
  \bibinfo {author} {\bibfnamefont {H.}~\bibnamefont {Xu}},\ and\ \bibinfo
  {author} {\bibfnamefont {S.}~\bibnamefont {Tarucha}},\ }\bibfield  {title}
  {\bibinfo {title} {{Dominant nonlocal superconducting proximity effect due to
  electron-electron interaction in a ballistic double nanowire}},\ }\href
  {https://doi.org/10.1126/sciadv.aaw2194} {\bibfield  {journal} {\bibinfo
  {journal} {Sci. Adv.}\ }\textbf {\bibinfo {volume} {5}},\ \bibinfo {pages}
  {eaaw2194} (\bibinfo {year} {2019})}\BibitemShut {NoStop}%
\bibitem [{\citenamefont {Kanne}\ \emph
  {et~al.}(2021{\natexlab{a}})\citenamefont {Kanne}, \citenamefont {Olsteins},
  \citenamefont {Marnauza}, \citenamefont {Vekris}, \citenamefont
  {Salda{\ifmmode\tilde{n}\else\~{n}\fi}a}, \citenamefont {Lori\`{c}},
  \citenamefont {Schlosser}, \citenamefont {Ross}, \citenamefont {Csonka},
  \citenamefont {Grove-Rasmussen},\ and\ \citenamefont
  {Nyg{\aa}rd}}]{Kanne2021Nov}%
  \BibitemOpen
  \bibfield  {author} {\bibinfo {author} {\bibfnamefont {T.}~\bibnamefont
  {Kanne}}, \bibinfo {author} {\bibfnamefont {D.}~\bibnamefont {Olsteins}},
  \bibinfo {author} {\bibfnamefont {M.}~\bibnamefont {Marnauza}}, \bibinfo
  {author} {\bibfnamefont {A.}~\bibnamefont {Vekris}}, \bibinfo {author}
  {\bibfnamefont {J.~C.~E.}\ \bibnamefont
  {Salda{\ifmmode\tilde{n}\else\~{n}\fi}a}}, \bibinfo {author} {\bibfnamefont
  {S.}~\bibnamefont {Lori\`{c}}}, \bibinfo {author} {\bibfnamefont {R.~D.}\
  \bibnamefont {Schlosser}}, \bibinfo {author} {\bibfnamefont {D.}~\bibnamefont
  {Ross}}, \bibinfo {author} {\bibfnamefont {S.}~\bibnamefont {Csonka}},
  \bibinfo {author} {\bibfnamefont {K.}~\bibnamefont {Grove-Rasmussen}},\ and\
  \bibinfo {author} {\bibfnamefont {J.}~\bibnamefont {Nyg{\aa}rd}},\ }\bibfield
   {title} {\bibinfo {title} {{Double Nanowires for Hybrid Quantum Devices}},\
  }\href {https://doi.org/10.1002/adfm.202107926} {\bibfield  {journal}
  {\bibinfo  {journal} {Adv. Funct. Mater.}\ }\textbf {\bibinfo {volume}
  {n/a}},\ \bibinfo {pages} {2107926} (\bibinfo {year}
  {2021}{\natexlab{a}})}\BibitemShut {NoStop}%
\bibitem [{\citenamefont {Vekris}\ \emph
  {et~al.}(2021{\natexlab{a}})\citenamefont {Vekris}, \citenamefont
  {Salda\~na}, \citenamefont {Kanne}, \citenamefont {Marnauza}, \citenamefont
  {Olsteins}, \citenamefont {Fan}, \citenamefont {Li}, \citenamefont
  {Hvid-Olsen}, \citenamefont {Qiu}, \citenamefont {Xu}, \citenamefont
  {Nyg\aa{}rd},\ and\ \citenamefont {Grove-Rasmussen}}]{VekrisJJ2021}%
  \BibitemOpen
  \bibfield  {author} {\bibinfo {author} {\bibfnamefont {A.}~\bibnamefont
  {Vekris}}, \bibinfo {author} {\bibfnamefont {J.~C.~E.}\ \bibnamefont
  {Salda\~na}}, \bibinfo {author} {\bibfnamefont {T.}~\bibnamefont {Kanne}},
  \bibinfo {author} {\bibfnamefont {M.}~\bibnamefont {Marnauza}}, \bibinfo
  {author} {\bibfnamefont {D.}~\bibnamefont {Olsteins}}, \bibinfo {author}
  {\bibfnamefont {F.}~\bibnamefont {Fan}}, \bibinfo {author} {\bibfnamefont
  {X.}~\bibnamefont {Li}}, \bibinfo {author} {\bibfnamefont {T.}~\bibnamefont
  {Hvid-Olsen}}, \bibinfo {author} {\bibfnamefont {X.}~\bibnamefont {Qiu}},
  \bibinfo {author} {\bibfnamefont {H.}~\bibnamefont {Xu}}, \bibinfo {author}
  {\bibfnamefont {J.}~\bibnamefont {Nyg\aa{}rd}},\ and\ \bibinfo {author}
  {\bibfnamefont {K.}~\bibnamefont {Grove-Rasmussen}},\ }\bibfield  {title}
  {\bibinfo {title} {Josephson junctions in double nanowires bridged by in-situ
  deposited superconductors},\ }\href
  {https://doi.org/10.1103/PhysRevResearch.3.033240} {\bibfield  {journal}
  {\bibinfo  {journal} {Phys. Rev. Research}\ }\textbf {\bibinfo {volume}
  {3}},\ \bibinfo {pages} {033240} (\bibinfo {year}
  {2021}{\natexlab{a}})}\BibitemShut {NoStop}%
\bibitem [{\citenamefont {Vekris}\ \emph
  {et~al.}(2021{\natexlab{b}})\citenamefont {Vekris}, \citenamefont
  {Estrada~Salda{\ifmmode\tilde{n}\else\~{n}\fi}a}, \citenamefont
  {de~Bruijckere}, \citenamefont {Lori{\ifmmode\acute{c}\else\'{c}\fi}},
  \citenamefont {Kanne}, \citenamefont {Marnauza}, \citenamefont {Olsteins},
  \citenamefont {Nyg{\aa}rd},\ and\ \citenamefont
  {Grove-Rasmussen}}]{Vekris2021Sep}%
  \BibitemOpen
  \bibfield  {author} {\bibinfo {author} {\bibfnamefont {A.}~\bibnamefont
  {Vekris}}, \bibinfo {author} {\bibfnamefont {J.~C.}\ \bibnamefont
  {Estrada~Salda{\ifmmode\tilde{n}\else\~{n}\fi}a}}, \bibinfo {author}
  {\bibfnamefont {J.}~\bibnamefont {de~Bruijckere}}, \bibinfo {author}
  {\bibfnamefont {S.}~\bibnamefont {Lori{\ifmmode\acute{c}\else\'{c}\fi}}},
  \bibinfo {author} {\bibfnamefont {T.}~\bibnamefont {Kanne}}, \bibinfo
  {author} {\bibfnamefont {M.}~\bibnamefont {Marnauza}}, \bibinfo {author}
  {\bibfnamefont {D.}~\bibnamefont {Olsteins}}, \bibinfo {author}
  {\bibfnamefont {J.}~\bibnamefont {Nyg{\aa}rd}},\ and\ \bibinfo {author}
  {\bibfnamefont {K.}~\bibnamefont {Grove-Rasmussen}},\ }\bibfield  {title}
  {\bibinfo {title} {{Asymmetric Little{\textendash}Parks oscillations in full
  shell double nanowires - Scientific Reports}},\ }\href
  {https://doi.org/10.1038/s41598-021-97780-9} {\bibfield  {journal} {\bibinfo
  {journal} {Sci. Rep.}\ }\textbf {\bibinfo {volume} {11}},\ \bibinfo {pages}
  {1} (\bibinfo {year} {2021}{\natexlab{b}})}\BibitemShut {NoStop}%
\bibitem [{\citenamefont {Terhal}\ \emph {et~al.}(2012)\citenamefont {Terhal},
  \citenamefont {Hassler},\ and\ \citenamefont {DiVincenzo}}]{Terhal2012PRL}%
  \BibitemOpen
  \bibfield  {author} {\bibinfo {author} {\bibfnamefont {B.~M.}\ \bibnamefont
  {Terhal}}, \bibinfo {author} {\bibfnamefont {F.}~\bibnamefont {Hassler}},\
  and\ \bibinfo {author} {\bibfnamefont {D.~P.}\ \bibnamefont {DiVincenzo}},\
  }\bibfield  {title} {\bibinfo {title} {From majorana fermions to topological
  order},\ }\href {https://doi.org/10.1103/PhysRevLett.108.260504} {\bibfield
  {journal} {\bibinfo  {journal} {Phys. Rev. Lett.}\ }\textbf {\bibinfo
  {volume} {108}},\ \bibinfo {pages} {260504} (\bibinfo {year}
  {2012})}\BibitemShut {NoStop}%
\bibitem [{\citenamefont {Plugge}\ \emph {et~al.}(2016)\citenamefont {Plugge},
  \citenamefont {Landau}, \citenamefont {Sela}, \citenamefont {Altland},
  \citenamefont {Flensberg},\ and\ \citenamefont {Egger}}]{Plugge2016PRB}%
  \BibitemOpen
  \bibfield  {author} {\bibinfo {author} {\bibfnamefont {S.}~\bibnamefont
  {Plugge}}, \bibinfo {author} {\bibfnamefont {L.~A.}\ \bibnamefont {Landau}},
  \bibinfo {author} {\bibfnamefont {E.}~\bibnamefont {Sela}}, \bibinfo {author}
  {\bibfnamefont {A.}~\bibnamefont {Altland}}, \bibinfo {author} {\bibfnamefont
  {K.}~\bibnamefont {Flensberg}},\ and\ \bibinfo {author} {\bibfnamefont
  {R.}~\bibnamefont {Egger}},\ }\bibfield  {title} {\bibinfo {title} {Roadmap
  to majorana surface codes},\ }\href
  {https://doi.org/10.1103/PhysRevB.94.174514} {\bibfield  {journal} {\bibinfo
  {journal} {Phys. Rev. B}\ }\textbf {\bibinfo {volume} {94}},\ \bibinfo
  {pages} {174514} (\bibinfo {year} {2016})}\BibitemShut {NoStop}%
\bibitem [{\citenamefont {Plugge}\ \emph {et~al.}(2017)\citenamefont {Plugge},
  \citenamefont {Rasmussen}, \citenamefont {Egger},\ and\ \citenamefont
  {Flensberg}}]{Plugge2017NJP}%
  \BibitemOpen
  \bibfield  {author} {\bibinfo {author} {\bibfnamefont {S.}~\bibnamefont
  {Plugge}}, \bibinfo {author} {\bibfnamefont {A.}~\bibnamefont {Rasmussen}},
  \bibinfo {author} {\bibfnamefont {R.}~\bibnamefont {Egger}},\ and\ \bibinfo
  {author} {\bibfnamefont {K.}~\bibnamefont {Flensberg}},\ }\bibfield  {title}
  {\bibinfo {title} {Majorana box qubits},\ }\href
  {https://doi.org/10.1088/1367-2630/aa54e1} {\bibfield  {journal} {\bibinfo
  {journal} {New J. Phys.}\ }\textbf {\bibinfo {volume} {19}},\ \bibinfo
  {pages} {012001} (\bibinfo {year} {2017})}\BibitemShut {NoStop}%
\bibitem [{\citenamefont {B{\ifmmode\acute{e}\else\'{e}\fi}ri}\ and\
  \citenamefont {Cooper}(2012)}]{Beri2012Oct}%
  \BibitemOpen
  \bibfield  {author} {\bibinfo {author} {\bibfnamefont {B.}~\bibnamefont
  {B{\ifmmode\acute{e}\else\'{e}\fi}ri}}\ and\ \bibinfo {author} {\bibfnamefont
  {N.~R.}\ \bibnamefont {Cooper}},\ }\bibfield  {title} {\bibinfo {title}
  {{Topological Kondo Effect with Majorana Fermions}},\ }\href
  {https://doi.org/10.1103/PhysRevLett.109.156803} {\bibfield  {journal}
  {\bibinfo  {journal} {Phys. Rev. Lett.}\ }\textbf {\bibinfo {volume} {109}},\
  \bibinfo {pages} {156803} (\bibinfo {year} {2012})}\BibitemShut {NoStop}%
\bibitem [{\citenamefont {B\'eri}(2013)}]{Beri2013}%
  \BibitemOpen
  \bibfield  {author} {\bibinfo {author} {\bibfnamefont {B.}~\bibnamefont
  {B\'eri}},\ }\bibfield  {title} {\bibinfo {title} {Majorana-klein
  hybridization in topological superconductor junctions},\ }\href
  {https://doi.org/10.1103/PhysRevLett.110.216803} {\bibfield  {journal}
  {\bibinfo  {journal} {Phys. Rev. Lett.}\ }\textbf {\bibinfo {volume} {110}},\
  \bibinfo {pages} {216803} (\bibinfo {year} {2013})}\BibitemShut {NoStop}%
\bibitem [{\citenamefont {Altland}\ and\ \citenamefont
  {Egger}(2013)}]{Altland2013May}%
  \BibitemOpen
  \bibfield  {author} {\bibinfo {author} {\bibfnamefont {A.}~\bibnamefont
  {Altland}}\ and\ \bibinfo {author} {\bibfnamefont {R.}~\bibnamefont
  {Egger}},\ }\bibfield  {title} {\bibinfo {title} {{Multiterminal
  Coulomb-Majorana Junction}},\ }\href
  {https://doi.org/10.1103/PhysRevLett.110.196401} {\bibfield  {journal}
  {\bibinfo  {journal} {Phys. Rev. Lett.}\ }\textbf {\bibinfo {volume} {110}},\
  \bibinfo {pages} {196401} (\bibinfo {year} {2013})}\BibitemShut {NoStop}%
\bibitem [{\citenamefont {Tuominen}\ \emph {et~al.}(1992)\citenamefont
  {Tuominen}, \citenamefont {Hergenrother}, \citenamefont {Tighe},\ and\
  \citenamefont {Tinkham}}]{PhysRevLett.69.1997}%
  \BibitemOpen
  \bibfield  {author} {\bibinfo {author} {\bibfnamefont {M.~T.}\ \bibnamefont
  {Tuominen}}, \bibinfo {author} {\bibfnamefont {J.~M.}\ \bibnamefont
  {Hergenrother}}, \bibinfo {author} {\bibfnamefont {T.~S.}\ \bibnamefont
  {Tighe}},\ and\ \bibinfo {author} {\bibfnamefont {M.}~\bibnamefont
  {Tinkham}},\ }\bibfield  {title} {\bibinfo {title} {Experimental evidence for
  parity-based 2e periodicity in a superconducting single-electron tunneling
  transistor},\ }\href {https://doi.org/10.1103/PhysRevLett.69.1997} {\bibfield
   {journal} {\bibinfo  {journal} {Phys. Rev. Lett.}\ }\textbf {\bibinfo
  {volume} {69}},\ \bibinfo {pages} {1997} (\bibinfo {year}
  {1992})}\BibitemShut {NoStop}%
\bibitem [{\citenamefont {Lafarge}\ \emph {et~al.}(1993)\citenamefont
  {Lafarge}, \citenamefont {Joyez}, \citenamefont {Esteve}, \citenamefont
  {Urbina},\ and\ \citenamefont {Devoret}}]{PhysRevLett.70.994}%
  \BibitemOpen
  \bibfield  {author} {\bibinfo {author} {\bibfnamefont {P.}~\bibnamefont
  {Lafarge}}, \bibinfo {author} {\bibfnamefont {P.}~\bibnamefont {Joyez}},
  \bibinfo {author} {\bibfnamefont {D.}~\bibnamefont {Esteve}}, \bibinfo
  {author} {\bibfnamefont {C.}~\bibnamefont {Urbina}},\ and\ \bibinfo {author}
  {\bibfnamefont {M.~H.}\ \bibnamefont {Devoret}},\ }\bibfield  {title}
  {\bibinfo {title} {Measurement of the even-odd free-energy difference of an
  isolated superconductor},\ }\href
  {https://doi.org/10.1103/PhysRevLett.70.994} {\bibfield  {journal} {\bibinfo
  {journal} {Phys. Rev. Lett.}\ }\textbf {\bibinfo {volume} {70}},\ \bibinfo
  {pages} {994} (\bibinfo {year} {1993})}\BibitemShut {NoStop}%
\bibitem [{\citenamefont {Joyez}\ \emph {et~al.}(1994)\citenamefont {Joyez},
  \citenamefont {Lafarge}, \citenamefont {Filipe}, \citenamefont {Esteve},\
  and\ \citenamefont {Devoret}}]{JoyezPRL1994}%
  \BibitemOpen
  \bibfield  {author} {\bibinfo {author} {\bibfnamefont {P.}~\bibnamefont
  {Joyez}}, \bibinfo {author} {\bibfnamefont {P.}~\bibnamefont {Lafarge}},
  \bibinfo {author} {\bibfnamefont {A.}~\bibnamefont {Filipe}}, \bibinfo
  {author} {\bibfnamefont {D.}~\bibnamefont {Esteve}},\ and\ \bibinfo {author}
  {\bibfnamefont {M.~H.}\ \bibnamefont {Devoret}},\ }\bibfield  {title}
  {\bibinfo {title} {Observation of parity-induced suppression of josephson
  tunneling in the superconducting single electron transistor},\ }\href
  {https://doi.org/10.1103/PhysRevLett.72.2458} {\bibfield  {journal} {\bibinfo
   {journal} {Phys. Rev. Lett.}\ }\textbf {\bibinfo {volume} {72}},\ \bibinfo
  {pages} {2458} (\bibinfo {year} {1994})}\BibitemShut {NoStop}%
\bibitem [{\citenamefont {Averin}\ and\ \citenamefont
  {Nazarov}(1992)}]{Averin1992Sep}%
  \BibitemOpen
  \bibfield  {author} {\bibinfo {author} {\bibfnamefont {D.~V.}\ \bibnamefont
  {Averin}}\ and\ \bibinfo {author} {\bibfnamefont {{\relax Yu}.~V.}\
  \bibnamefont {Nazarov}},\ }\bibfield  {title} {\bibinfo {title}
  {{Single-electron charging of a superconducting island}},\ }\href
  {https://doi.org/10.1103/PhysRevLett.69.1993} {\bibfield  {journal} {\bibinfo
   {journal} {Phys. Rev. Lett.}\ }\textbf {\bibinfo {volume} {69}},\ \bibinfo
  {pages} {1993} (\bibinfo {year} {1992})}\BibitemShut {NoStop}%
\bibitem [{\citenamefont {Eiles}\ \emph {et~al.}(1993)\citenamefont {Eiles},
  \citenamefont {Martinis},\ and\ \citenamefont {Devoret}}]{Eiles1993Mar}%
  \BibitemOpen
  \bibfield  {author} {\bibinfo {author} {\bibfnamefont {T.~M.}\ \bibnamefont
  {Eiles}}, \bibinfo {author} {\bibfnamefont {J.~M.}\ \bibnamefont
  {Martinis}},\ and\ \bibinfo {author} {\bibfnamefont {M.~H.}\ \bibnamefont
  {Devoret}},\ }\bibfield  {title} {\bibinfo {title} {{Even-odd asymmetry of a
  superconductor revealed by the Coulomb blockade of Andreev reflection}},\
  }\href {https://doi.org/10.1103/PhysRevLett.70.1862} {\bibfield  {journal}
  {\bibinfo  {journal} {Phys. Rev. Lett.}\ }\textbf {\bibinfo {volume} {70}},\
  \bibinfo {pages} {1862} (\bibinfo {year} {1993})}\BibitemShut {NoStop}%
\bibitem [{\citenamefont {Higginbotham}\ \emph {et~al.}(2015)\citenamefont
  {Higginbotham}, \citenamefont {Albrecht}, \citenamefont {Kir{\v{s}}anskas},
  \citenamefont {Chang}, \citenamefont {Kuemmeth}, \citenamefont {Krogstrup},
  \citenamefont {Jespersen}, \citenamefont {Nyg{\aa}rd}, \citenamefont
  {Flensberg},\ and\ \citenamefont {Marcus}}]{higginbotham2015parity}%
  \BibitemOpen
  \bibfield  {author} {\bibinfo {author} {\bibfnamefont {A.~P.}\ \bibnamefont
  {Higginbotham}}, \bibinfo {author} {\bibfnamefont {S.~M.}\ \bibnamefont
  {Albrecht}}, \bibinfo {author} {\bibfnamefont {G.}~\bibnamefont
  {Kir{\v{s}}anskas}}, \bibinfo {author} {\bibfnamefont {W.}~\bibnamefont
  {Chang}}, \bibinfo {author} {\bibfnamefont {F.}~\bibnamefont {Kuemmeth}},
  \bibinfo {author} {\bibfnamefont {P.}~\bibnamefont {Krogstrup}}, \bibinfo
  {author} {\bibfnamefont {T.~S.}\ \bibnamefont {Jespersen}}, \bibinfo {author}
  {\bibfnamefont {J.}~\bibnamefont {Nyg{\aa}rd}}, \bibinfo {author}
  {\bibfnamefont {K.}~\bibnamefont {Flensberg}},\ and\ \bibinfo {author}
  {\bibfnamefont {C.~M.}\ \bibnamefont {Marcus}},\ }\bibfield  {title}
  {\bibinfo {title} {Parity lifetime of bound states in a proximitized
  semiconductor nanowire},\ }\href {https://doi.org/10.1038/nphys3461}
  {\bibfield  {journal} {\bibinfo  {journal} {Nature Physics}\ }\textbf
  {\bibinfo {volume} {11}},\ \bibinfo {pages} {1017} (\bibinfo {year}
  {2015})}\BibitemShut {NoStop}%
\bibitem [{\citenamefont {Albrecht}\ \emph {et~al.}(2017)\citenamefont
  {Albrecht}, \citenamefont {Hansen}, \citenamefont {Higginbotham},
  \citenamefont {Kuemmeth}, \citenamefont {Jespersen}, \citenamefont
  {Nyg{\aa}rd}, \citenamefont {Krogstrup}, \citenamefont {Danon}, \citenamefont
  {Flensberg},\ and\ \citenamefont {Marcus}}]{Albrecht2017Mar}%
  \BibitemOpen
  \bibfield  {author} {\bibinfo {author} {\bibfnamefont {S.~M.}\ \bibnamefont
  {Albrecht}}, \bibinfo {author} {\bibfnamefont {E.~B.}\ \bibnamefont
  {Hansen}}, \bibinfo {author} {\bibfnamefont {A.~P.}\ \bibnamefont
  {Higginbotham}}, \bibinfo {author} {\bibfnamefont {F.}~\bibnamefont
  {Kuemmeth}}, \bibinfo {author} {\bibfnamefont {T.~S.}\ \bibnamefont
  {Jespersen}}, \bibinfo {author} {\bibfnamefont {J.}~\bibnamefont
  {Nyg{\aa}rd}}, \bibinfo {author} {\bibfnamefont {P.}~\bibnamefont
  {Krogstrup}}, \bibinfo {author} {\bibfnamefont {J.}~\bibnamefont {Danon}},
  \bibinfo {author} {\bibfnamefont {K.}~\bibnamefont {Flensberg}},\ and\
  \bibinfo {author} {\bibfnamefont {C.~M.}\ \bibnamefont {Marcus}},\ }\bibfield
   {title} {\bibinfo {title} {{Transport Signatures of Quasiparticle Poisoning
  in a Majorana Island}},\ }\href
  {https://doi.org/10.1103/PhysRevLett.118.137701} {\bibfield  {journal}
  {\bibinfo  {journal} {Phys. Rev. Lett.}\ }\textbf {\bibinfo {volume} {118}},\
  \bibinfo {pages} {137701} (\bibinfo {year} {2017})}\BibitemShut {NoStop}%
\bibitem [{\citenamefont {M\'enard}\ \emph {et~al.}(2019)\citenamefont
  {M\'enard}, \citenamefont {Malinowski}, \citenamefont {Puglia}, \citenamefont
  {Pikulin}, \citenamefont {Karzig}, \citenamefont {Bauer}, \citenamefont
  {Krogstrup},\ and\ \citenamefont {Marcus}}]{MenardPRB2019}%
  \BibitemOpen
  \bibfield  {author} {\bibinfo {author} {\bibfnamefont {G.~C.}\ \bibnamefont
  {M\'enard}}, \bibinfo {author} {\bibfnamefont {F.~K.}\ \bibnamefont
  {Malinowski}}, \bibinfo {author} {\bibfnamefont {D.}~\bibnamefont {Puglia}},
  \bibinfo {author} {\bibfnamefont {D.~I.}\ \bibnamefont {Pikulin}}, \bibinfo
  {author} {\bibfnamefont {T.}~\bibnamefont {Karzig}}, \bibinfo {author}
  {\bibfnamefont {B.}~\bibnamefont {Bauer}}, \bibinfo {author} {\bibfnamefont
  {P.}~\bibnamefont {Krogstrup}},\ and\ \bibinfo {author} {\bibfnamefont
  {C.~M.}\ \bibnamefont {Marcus}},\ }\bibfield  {title} {\bibinfo {title}
  {Suppressing quasiparticle poisoning with a voltage-controlled filter},\
  }\href {https://doi.org/10.1103/PhysRevB.100.165307} {\bibfield  {journal}
  {\bibinfo  {journal} {Phys. Rev. B}\ }\textbf {\bibinfo {volume} {100}},\
  \bibinfo {pages} {165307} (\bibinfo {year} {2019})}\BibitemShut {NoStop}%
\bibitem [{\citenamefont {Albrecht}\ \emph {et~al.}(2016)\citenamefont
  {Albrecht}, \citenamefont {Higginbotham}, \citenamefont {Madsen},
  \citenamefont {Kuemmeth}, \citenamefont {Jespersen}, \citenamefont
  {Nyg{\aa}rd}, \citenamefont {Krogstrup},\ and\ \citenamefont
  {Marcus}}]{Albrecht2016Mar}%
  \BibitemOpen
  \bibfield  {author} {\bibinfo {author} {\bibfnamefont {S.~M.}\ \bibnamefont
  {Albrecht}}, \bibinfo {author} {\bibfnamefont {A.~P.}\ \bibnamefont
  {Higginbotham}}, \bibinfo {author} {\bibfnamefont {M.}~\bibnamefont
  {Madsen}}, \bibinfo {author} {\bibfnamefont {F.}~\bibnamefont {Kuemmeth}},
  \bibinfo {author} {\bibfnamefont {T.~S.}\ \bibnamefont {Jespersen}}, \bibinfo
  {author} {\bibfnamefont {J.}~\bibnamefont {Nyg{\aa}rd}}, \bibinfo {author}
  {\bibfnamefont {P.}~\bibnamefont {Krogstrup}},\ and\ \bibinfo {author}
  {\bibfnamefont {C.~M.}\ \bibnamefont {Marcus}},\ }\bibfield  {title}
  {\bibinfo {title} {{Exponential protection of zero modes in Majorana
  islands}},\ }\href {https://doi.org/10.1038/nature17162} {\bibfield
  {journal} {\bibinfo  {journal} {Nature}\ }\textbf {\bibinfo {volume} {531}},\
  \bibinfo {pages} {206} (\bibinfo {year} {2016})}\BibitemShut {NoStop}%
\bibitem [{\citenamefont {Sherman}\ \emph {et~al.}(2017)\citenamefont
  {Sherman}, \citenamefont {Yodh}, \citenamefont {Albrecht}, \citenamefont
  {Nyg{\aa}rd}, \citenamefont {Krogstrup},\ and\ \citenamefont
  {Marcus}}]{Sherman2017Mar}%
  \BibitemOpen
  \bibfield  {author} {\bibinfo {author} {\bibfnamefont {D.}~\bibnamefont
  {Sherman}}, \bibinfo {author} {\bibfnamefont {J.~S.}\ \bibnamefont {Yodh}},
  \bibinfo {author} {\bibfnamefont {S.~M.}\ \bibnamefont {Albrecht}}, \bibinfo
  {author} {\bibfnamefont {J.}~\bibnamefont {Nyg{\aa}rd}}, \bibinfo {author}
  {\bibfnamefont {P.}~\bibnamefont {Krogstrup}},\ and\ \bibinfo {author}
  {\bibfnamefont {C.~M.}\ \bibnamefont {Marcus}},\ }\bibfield  {title}
  {\bibinfo {title} {{Normal, superconducting and topological regimes of hybrid
  double quantum dots - Nature Nanotechnology}},\ }\href
  {https://doi.org/10.1038/nnano.2016.227} {\bibfield  {journal} {\bibinfo
  {journal} {Nat. Nanotechnol.}\ }\textbf {\bibinfo {volume} {12}},\ \bibinfo
  {pages} {212} (\bibinfo {year} {2017})}\BibitemShut {NoStop}%
\bibitem [{\citenamefont {Shen}\ \emph {et~al.}(2018)\citenamefont {Shen},
  \citenamefont {Heedt}, \citenamefont {Borsoi}, \citenamefont {van Heck},
  \citenamefont {Gazibegovic}, \citenamefont {Op~het Veld}, \citenamefont
  {Car}, \citenamefont {Logan}, \citenamefont {Pendharkar}, \citenamefont
  {Ramakers}, \citenamefont {Wang}, \citenamefont {Xu}, \citenamefont {Bouman},
  \citenamefont {Geresdi}, \citenamefont {Palmstr{\o}m}, \citenamefont
  {Bakkers},\ and\ \citenamefont {Kouwenhoven}}]{Shen2018Nov}%
  \BibitemOpen
  \bibfield  {author} {\bibinfo {author} {\bibfnamefont {J.}~\bibnamefont
  {Shen}}, \bibinfo {author} {\bibfnamefont {S.}~\bibnamefont {Heedt}},
  \bibinfo {author} {\bibfnamefont {F.}~\bibnamefont {Borsoi}}, \bibinfo
  {author} {\bibfnamefont {B.}~\bibnamefont {van Heck}}, \bibinfo {author}
  {\bibfnamefont {S.}~\bibnamefont {Gazibegovic}}, \bibinfo {author}
  {\bibfnamefont {R.~L.~M.}\ \bibnamefont {Op~het Veld}}, \bibinfo {author}
  {\bibfnamefont {D.}~\bibnamefont {Car}}, \bibinfo {author} {\bibfnamefont
  {J.~A.}\ \bibnamefont {Logan}}, \bibinfo {author} {\bibfnamefont
  {M.}~\bibnamefont {Pendharkar}}, \bibinfo {author} {\bibfnamefont {S.~J.~J.}\
  \bibnamefont {Ramakers}}, \bibinfo {author} {\bibfnamefont {G.}~\bibnamefont
  {Wang}}, \bibinfo {author} {\bibfnamefont {D.}~\bibnamefont {Xu}}, \bibinfo
  {author} {\bibfnamefont {D.}~\bibnamefont {Bouman}}, \bibinfo {author}
  {\bibfnamefont {A.}~\bibnamefont {Geresdi}}, \bibinfo {author} {\bibfnamefont
  {C.~J.}\ \bibnamefont {Palmstr{\o}m}}, \bibinfo {author} {\bibfnamefont
  {E.~P. A.~M.}\ \bibnamefont {Bakkers}},\ and\ \bibinfo {author}
  {\bibfnamefont {L.~P.}\ \bibnamefont {Kouwenhoven}},\ }\bibfield  {title}
  {\bibinfo {title} {{Parity transitions in the superconducting ground state of
  hybrid InSb{\textendash}Al Coulomb islands - Nature Communications}},\ }\href
  {https://doi.org/10.1038/s41467-018-07279-7} {\bibfield  {journal} {\bibinfo
  {journal} {Nat. Commun.}\ }\textbf {\bibinfo {volume} {9}},\ \bibinfo {pages}
  {1} (\bibinfo {year} {2018})}\BibitemShut {NoStop}%
\bibitem [{\citenamefont {Sestoft}\ \emph {et~al.}(2018)\citenamefont
  {Sestoft}, \citenamefont {Kanne}, \citenamefont {Gejl}, \citenamefont {von
  Soosten}, \citenamefont {Yodh}, \citenamefont {Sherman}, \citenamefont
  {Tarasinski}, \citenamefont {Wimmer}, \citenamefont {Johnson}, \citenamefont
  {Deng}, \citenamefont {Nyg\aa{}rd}, \citenamefont {Jespersen}, \citenamefont
  {Marcus},\ and\ \citenamefont {Krogstrup}}]{SestoftPRM2018}%
  \BibitemOpen
  \bibfield  {author} {\bibinfo {author} {\bibfnamefont {J.~E.}\ \bibnamefont
  {Sestoft}}, \bibinfo {author} {\bibfnamefont {T.}~\bibnamefont {Kanne}},
  \bibinfo {author} {\bibfnamefont {A.~N.}\ \bibnamefont {Gejl}}, \bibinfo
  {author} {\bibfnamefont {M.}~\bibnamefont {von Soosten}}, \bibinfo {author}
  {\bibfnamefont {J.~S.}\ \bibnamefont {Yodh}}, \bibinfo {author}
  {\bibfnamefont {D.}~\bibnamefont {Sherman}}, \bibinfo {author} {\bibfnamefont
  {B.}~\bibnamefont {Tarasinski}}, \bibinfo {author} {\bibfnamefont
  {M.}~\bibnamefont {Wimmer}}, \bibinfo {author} {\bibfnamefont
  {E.}~\bibnamefont {Johnson}}, \bibinfo {author} {\bibfnamefont
  {M.}~\bibnamefont {Deng}}, \bibinfo {author} {\bibfnamefont {J.}~\bibnamefont
  {Nyg\aa{}rd}}, \bibinfo {author} {\bibfnamefont {T.~S.}\ \bibnamefont
  {Jespersen}}, \bibinfo {author} {\bibfnamefont {C.~M.}\ \bibnamefont
  {Marcus}},\ and\ \bibinfo {author} {\bibfnamefont {P.}~\bibnamefont
  {Krogstrup}},\ }\bibfield  {title} {\bibinfo {title} {Engineering hybrid
  epitaxial inassb/al nanowires for stronger topological protection},\ }\href
  {https://doi.org/10.1103/PhysRevMaterials.2.044202} {\bibfield  {journal}
  {\bibinfo  {journal} {Phys. Rev. Materials}\ }\textbf {\bibinfo {volume}
  {2}},\ \bibinfo {pages} {044202} (\bibinfo {year} {2018})}\BibitemShut
  {NoStop}%
\bibitem [{\citenamefont {O'Farrell}\ \emph {et~al.}(2018)\citenamefont
  {O'Farrell}, \citenamefont {Drachmann}, \citenamefont {Hell}, \citenamefont
  {Fornieri}, \citenamefont {Whiticar}, \citenamefont {Hansen}, \citenamefont
  {Gronin}, \citenamefont {Gardner}, \citenamefont {Thomas}, \citenamefont
  {Manfra}, \citenamefont {Flensberg}, \citenamefont {Marcus},\ and\
  \citenamefont {Nichele}}]{OfarrellPRL2018}%
  \BibitemOpen
  \bibfield  {author} {\bibinfo {author} {\bibfnamefont {E.~C.~T.}\
  \bibnamefont {O'Farrell}}, \bibinfo {author} {\bibfnamefont {A.~C.~C.}\
  \bibnamefont {Drachmann}}, \bibinfo {author} {\bibfnamefont {M.}~\bibnamefont
  {Hell}}, \bibinfo {author} {\bibfnamefont {A.}~\bibnamefont {Fornieri}},
  \bibinfo {author} {\bibfnamefont {A.~M.}\ \bibnamefont {Whiticar}}, \bibinfo
  {author} {\bibfnamefont {E.~B.}\ \bibnamefont {Hansen}}, \bibinfo {author}
  {\bibfnamefont {S.}~\bibnamefont {Gronin}}, \bibinfo {author} {\bibfnamefont
  {G.~C.}\ \bibnamefont {Gardner}}, \bibinfo {author} {\bibfnamefont
  {C.}~\bibnamefont {Thomas}}, \bibinfo {author} {\bibfnamefont {M.~J.}\
  \bibnamefont {Manfra}}, \bibinfo {author} {\bibfnamefont {K.}~\bibnamefont
  {Flensberg}}, \bibinfo {author} {\bibfnamefont {C.~M.}\ \bibnamefont
  {Marcus}},\ and\ \bibinfo {author} {\bibfnamefont {F.}~\bibnamefont
  {Nichele}},\ }\bibfield  {title} {\bibinfo {title} {Hybridization of subgap
  states in one-dimensional superconductor-semiconductor coulomb islands},\
  }\href {https://doi.org/10.1103/PhysRevLett.121.256803} {\bibfield  {journal}
  {\bibinfo  {journal} {Phys. Rev. Lett.}\ }\textbf {\bibinfo {volume} {121}},\
  \bibinfo {pages} {256803} (\bibinfo {year} {2018})}\BibitemShut {NoStop}%
\bibitem [{\citenamefont {van Veen}\ \emph {et~al.}(2018)\citenamefont {van
  Veen}, \citenamefont {Proutski}, \citenamefont {Karzig}, \citenamefont
  {Pikulin}, \citenamefont {Lutchyn}, \citenamefont {Nyg\aa{}rd}, \citenamefont
  {Krogstrup}, \citenamefont {Geresdi}, \citenamefont {Kouwenhoven},\ and\
  \citenamefont {Watson}}]{vanVeenPRB2018}%
  \BibitemOpen
  \bibfield  {author} {\bibinfo {author} {\bibfnamefont {J.}~\bibnamefont {van
  Veen}}, \bibinfo {author} {\bibfnamefont {A.}~\bibnamefont {Proutski}},
  \bibinfo {author} {\bibfnamefont {T.}~\bibnamefont {Karzig}}, \bibinfo
  {author} {\bibfnamefont {D.~I.}\ \bibnamefont {Pikulin}}, \bibinfo {author}
  {\bibfnamefont {R.~M.}\ \bibnamefont {Lutchyn}}, \bibinfo {author}
  {\bibfnamefont {J.}~\bibnamefont {Nyg\aa{}rd}}, \bibinfo {author}
  {\bibfnamefont {P.}~\bibnamefont {Krogstrup}}, \bibinfo {author}
  {\bibfnamefont {A.}~\bibnamefont {Geresdi}}, \bibinfo {author} {\bibfnamefont
  {L.~P.}\ \bibnamefont {Kouwenhoven}},\ and\ \bibinfo {author} {\bibfnamefont
  {J.~D.}\ \bibnamefont {Watson}},\ }\bibfield  {title} {\bibinfo {title}
  {Magnetic-field-dependent quasiparticle dynamics of nanowire
  single-cooper-pair transistors},\ }\href
  {https://doi.org/10.1103/PhysRevB.98.174502} {\bibfield  {journal} {\bibinfo
  {journal} {Phys. Rev. B}\ }\textbf {\bibinfo {volume} {98}},\ \bibinfo
  {pages} {174502} (\bibinfo {year} {2018})}\BibitemShut {NoStop}%
\bibitem [{\citenamefont {Carrad}\ \emph {et~al.}(2020)\citenamefont {Carrad},
  \citenamefont {Bjergfelt}, \citenamefont {Kanne}, \citenamefont {Aagesen},
  \citenamefont {Krizek}, \citenamefont {Fiordaliso}, \citenamefont {Johnson},
  \citenamefont {Nygard},\ and\ \citenamefont {Jespersen}}]{CarradAdvMat2020}%
  \BibitemOpen
  \bibfield  {author} {\bibinfo {author} {\bibfnamefont {D.~J.}\ \bibnamefont
  {Carrad}}, \bibinfo {author} {\bibfnamefont {M.}~\bibnamefont {Bjergfelt}},
  \bibinfo {author} {\bibfnamefont {T.}~\bibnamefont {Kanne}}, \bibinfo
  {author} {\bibfnamefont {M.}~\bibnamefont {Aagesen}}, \bibinfo {author}
  {\bibfnamefont {F.}~\bibnamefont {Krizek}}, \bibinfo {author} {\bibfnamefont
  {E.~M.}\ \bibnamefont {Fiordaliso}}, \bibinfo {author} {\bibfnamefont
  {E.}~\bibnamefont {Johnson}}, \bibinfo {author} {\bibfnamefont
  {J.}~\bibnamefont {Nygard}},\ and\ \bibinfo {author} {\bibfnamefont {T.~S.}\
  \bibnamefont {Jespersen}},\ }\bibfield  {title} {\bibinfo {title} {{Shadow
  Epitaxy for In Situ Growth of Generic Semiconductor/Superconductor
  Hybrids}},\ }\bibfield  {journal} {\bibinfo  {journal} {{ADVANCED
  MATERIALS}}\ }\textbf {\bibinfo {volume} {{32}}},\ \href
  {https://doi.org/{10.1002/adma.201908411}} {{10.1002/adma.201908411}}
  (\bibinfo {year} {{2020}})\BibitemShut {NoStop}%
\bibitem [{\citenamefont {Shen}\ \emph {et~al.}(2021)\citenamefont {Shen},
  \citenamefont {Winkler}, \citenamefont {Borsoi}, \citenamefont {Heedt},
  \citenamefont {Levajac}, \citenamefont {Wang}, \citenamefont {van Driel},
  \citenamefont {Bouman}, \citenamefont {Gazibegovic}, \citenamefont
  {Op~Het~Veld}, \citenamefont {Car}, \citenamefont {Logan}, \citenamefont
  {Pendharkar}, \citenamefont {Palmstr\o{}m}, \citenamefont {Bakkers},
  \citenamefont {Kouwenhoven},\ and\ \citenamefont {van Heck}}]{Shen2021}%
  \BibitemOpen
  \bibfield  {author} {\bibinfo {author} {\bibfnamefont {J.}~\bibnamefont
  {Shen}}, \bibinfo {author} {\bibfnamefont {G.~W.}\ \bibnamefont {Winkler}},
  \bibinfo {author} {\bibfnamefont {F.}~\bibnamefont {Borsoi}}, \bibinfo
  {author} {\bibfnamefont {S.}~\bibnamefont {Heedt}}, \bibinfo {author}
  {\bibfnamefont {V.}~\bibnamefont {Levajac}}, \bibinfo {author} {\bibfnamefont
  {J.-Y.}\ \bibnamefont {Wang}}, \bibinfo {author} {\bibfnamefont
  {D.}~\bibnamefont {van Driel}}, \bibinfo {author} {\bibfnamefont
  {D.}~\bibnamefont {Bouman}}, \bibinfo {author} {\bibfnamefont
  {S.}~\bibnamefont {Gazibegovic}}, \bibinfo {author} {\bibfnamefont
  {R.~L.~M.}\ \bibnamefont {Op~Het~Veld}}, \bibinfo {author} {\bibfnamefont
  {D.}~\bibnamefont {Car}}, \bibinfo {author} {\bibfnamefont {J.~A.}\
  \bibnamefont {Logan}}, \bibinfo {author} {\bibfnamefont {M.}~\bibnamefont
  {Pendharkar}}, \bibinfo {author} {\bibfnamefont {C.~J.}\ \bibnamefont
  {Palmstr\o{}m}}, \bibinfo {author} {\bibfnamefont {E.~P. A.~M.}\ \bibnamefont
  {Bakkers}}, \bibinfo {author} {\bibfnamefont {L.~P.}\ \bibnamefont
  {Kouwenhoven}},\ and\ \bibinfo {author} {\bibfnamefont {B.}~\bibnamefont {van
  Heck}},\ }\bibfield  {title} {\bibinfo {title} {Full parity phase diagram of
  a proximitized nanowire island},\ }\href
  {https://doi.org/10.1103/PhysRevB.104.045422} {\bibfield  {journal} {\bibinfo
   {journal} {Phys. Rev. B}\ }\textbf {\bibinfo {volume} {104}},\ \bibinfo
  {pages} {045422} (\bibinfo {year} {2021})}\BibitemShut {NoStop}%
\bibitem [{\citenamefont {Fleckenstein}\ \emph {et~al.}(2018)\citenamefont
  {Fleckenstein}, \citenamefont {Dom\'{\i}nguez}, \citenamefont
  {Traverso~Ziani},\ and\ \citenamefont {Trauzettel}}]{Fleckenstein2018}%
  \BibitemOpen
  \bibfield  {author} {\bibinfo {author} {\bibfnamefont {C.}~\bibnamefont
  {Fleckenstein}}, \bibinfo {author} {\bibfnamefont {F.}~\bibnamefont
  {Dom\'{\i}nguez}}, \bibinfo {author} {\bibfnamefont {N.}~\bibnamefont
  {Traverso~Ziani}},\ and\ \bibinfo {author} {\bibfnamefont {B.}~\bibnamefont
  {Trauzettel}},\ }\bibfield  {title} {\bibinfo {title} {Decaying spectral
  oscillations in a majorana wire with finite coherence length},\ }\href
  {https://doi.org/10.1103/PhysRevB.97.155425} {\bibfield  {journal} {\bibinfo
  {journal} {Phys. Rev. B}\ }\textbf {\bibinfo {volume} {97}},\ \bibinfo
  {pages} {155425} (\bibinfo {year} {2018})}\BibitemShut {NoStop}%
\bibitem [{\citenamefont {Estrada~Salda{\ifmmode\tilde{n}\else\~{n}\fi}a}\
  \emph {et~al.}(2021)\citenamefont
  {Estrada~Salda{\ifmmode\tilde{n}\else\~{n}\fi}a}, \citenamefont {Vekris},
  \citenamefont
  {Pave{\ifmmode\check{s}\else\v{s}\fi}i{\ifmmode\check{c}\else\v{c}\fi}},
  \citenamefont {Krogstrup}, \citenamefont
  {{\ifmmode\check{Z}\else\v{Z}\fi}itko}, \citenamefont {Grove-Rasmussen},\
  and\ \citenamefont {Nyg{\aa}rd}}]{Saldana2022Jan}%
  \BibitemOpen
  \bibfield  {author} {\bibinfo {author} {\bibfnamefont {J.~C.}\ \bibnamefont
  {Estrada~Salda{\ifmmode\tilde{n}\else\~{n}\fi}a}}, \bibinfo {author}
  {\bibfnamefont {A.}~\bibnamefont {Vekris}}, \bibinfo {author} {\bibfnamefont
  {L.}~\bibnamefont
  {Pave{\ifmmode\check{s}\else\v{s}\fi}i{\ifmmode\check{c}\else\v{c}\fi}}},
  \bibinfo {author} {\bibfnamefont {P.}~\bibnamefont {Krogstrup}}, \bibinfo
  {author} {\bibfnamefont {R.}~\bibnamefont
  {{\ifmmode\check{Z}\else\v{Z}\fi}itko}}, \bibinfo {author} {\bibfnamefont
  {K.}~\bibnamefont {Grove-Rasmussen}},\ and\ \bibinfo {author} {\bibfnamefont
  {J.}~\bibnamefont {Nyg{\aa}rd}},\ }\bibfield  {title} {\bibinfo {title}
  {{Excitations in a superconducting Coulombic energy gap}},\ }\href
  {https://arxiv.org/abs/2101.10794v3} {\bibfield  {journal} {\bibinfo
  {journal} {arXiv}\ } (\bibinfo {year} {2021})},\ \Eprint
  {https://arxiv.org/abs/2101.10794} {2101.10794} \BibitemShut {NoStop}%
\bibitem [{\citenamefont {van Heck}\ \emph {et~al.}(2016)\citenamefont {van
  Heck}, \citenamefont {Lutchyn},\ and\ \citenamefont
  {Glazman}}]{vanHeckPRB2016}%
  \BibitemOpen
  \bibfield  {author} {\bibinfo {author} {\bibfnamefont {B.}~\bibnamefont {van
  Heck}}, \bibinfo {author} {\bibfnamefont {R.~M.}\ \bibnamefont {Lutchyn}},\
  and\ \bibinfo {author} {\bibfnamefont {L.~I.}\ \bibnamefont {Glazman}},\
  }\bibfield  {title} {\bibinfo {title} {Conductance of a proximitized nanowire
  in the coulomb blockade regime},\ }\href
  {https://doi.org/10.1103/PhysRevB.93.235431} {\bibfield  {journal} {\bibinfo
  {journal} {Phys. Rev. B}\ }\textbf {\bibinfo {volume} {93}},\ \bibinfo
  {pages} {235431} (\bibinfo {year} {2016})}\BibitemShut {NoStop}%
\bibitem [{\citenamefont {Vaitiek{\ifmmode\dot{e}\else\.{e}\fi}nas}\ \emph
  {et~al.}(2020)\citenamefont {Vaitiek{\ifmmode\dot{e}\else\.{e}\fi}nas},
  \citenamefont {Winkler}, \citenamefont {van Heck}, \citenamefont {Karzig},
  \citenamefont {Deng}, \citenamefont {Flensberg}, \citenamefont {Glazman},
  \citenamefont {Nayak}, \citenamefont {Krogstrup}, \citenamefont {Lutchyn},\
  and\ \citenamefont {Marcus}}]{Vaitiekenas2020Mar}%
  \BibitemOpen
  \bibfield  {author} {\bibinfo {author} {\bibfnamefont {S.}~\bibnamefont
  {Vaitiek{\ifmmode\dot{e}\else\.{e}\fi}nas}}, \bibinfo {author} {\bibfnamefont
  {G.~W.}\ \bibnamefont {Winkler}}, \bibinfo {author} {\bibfnamefont
  {B.}~\bibnamefont {van Heck}}, \bibinfo {author} {\bibfnamefont
  {T.}~\bibnamefont {Karzig}}, \bibinfo {author} {\bibfnamefont {M.-T.}\
  \bibnamefont {Deng}}, \bibinfo {author} {\bibfnamefont {K.}~\bibnamefont
  {Flensberg}}, \bibinfo {author} {\bibfnamefont {L.~I.}\ \bibnamefont
  {Glazman}}, \bibinfo {author} {\bibfnamefont {C.}~\bibnamefont {Nayak}},
  \bibinfo {author} {\bibfnamefont {P.}~\bibnamefont {Krogstrup}}, \bibinfo
  {author} {\bibfnamefont {R.~M.}\ \bibnamefont {Lutchyn}},\ and\ \bibinfo
  {author} {\bibfnamefont {C.~M.}\ \bibnamefont {Marcus}},\ }\bibfield  {title}
  {\bibinfo {title} {{Flux-induced topological superconductivity in full-shell
  nanowires}},\ }\href {https://doi.org/10.1126/science.aav3392} {\bibfield
  {journal} {\bibinfo  {journal} {Science}\ }\textbf {\bibinfo {volume}
  {367}},\ \bibinfo {pages} {eaav3392} (\bibinfo {year} {2020})}\BibitemShut
  {NoStop}%
\bibitem [{\citenamefont {Whiticar}\ \emph {et~al.}(2020)\citenamefont
  {Whiticar}, \citenamefont {Fornieri}, \citenamefont {O{'}Farrell},
  \citenamefont {Drachmann}, \citenamefont {Wang}, \citenamefont {Thomas},
  \citenamefont {Gronin}, \citenamefont {Kallaher}, \citenamefont {Gardner},
  \citenamefont {Manfra}, \citenamefont {Marcus},\ and\ \citenamefont
  {Nichele}}]{Whiticar2020Jun}%
  \BibitemOpen
  \bibfield  {author} {\bibinfo {author} {\bibfnamefont {A.~M.}\ \bibnamefont
  {Whiticar}}, \bibinfo {author} {\bibfnamefont {A.}~\bibnamefont {Fornieri}},
  \bibinfo {author} {\bibfnamefont {E.~C.~T.}\ \bibnamefont {O{'}Farrell}},
  \bibinfo {author} {\bibfnamefont {A.~C.~C.}\ \bibnamefont {Drachmann}},
  \bibinfo {author} {\bibfnamefont {T.}~\bibnamefont {Wang}}, \bibinfo {author}
  {\bibfnamefont {C.}~\bibnamefont {Thomas}}, \bibinfo {author} {\bibfnamefont
  {S.}~\bibnamefont {Gronin}}, \bibinfo {author} {\bibfnamefont
  {R.}~\bibnamefont {Kallaher}}, \bibinfo {author} {\bibfnamefont {G.~C.}\
  \bibnamefont {Gardner}}, \bibinfo {author} {\bibfnamefont {M.~J.}\
  \bibnamefont {Manfra}}, \bibinfo {author} {\bibfnamefont {C.~M.}\
  \bibnamefont {Marcus}},\ and\ \bibinfo {author} {\bibfnamefont
  {F.}~\bibnamefont {Nichele}},\ }\bibfield  {title} {\bibinfo {title}
  {{Coherent transport through a Majorana island in an
  Aharonov{\textendash}Bohm interferometer - Nature Communications}},\ }\href
  {https://doi.org/10.1038/s41467-020-16988-x} {\bibfield  {journal} {\bibinfo
  {journal} {Nat. Commun.}\ }\textbf {\bibinfo {volume} {11}},\ \bibinfo
  {pages} {1} (\bibinfo {year} {2020})}\BibitemShut {NoStop}%
\bibitem [{\citenamefont {Razmadze}\ \emph {et~al.}(2019)\citenamefont
  {Razmadze}, \citenamefont {Sabonis}, \citenamefont {Malinowski},
  \citenamefont {M\'enard}, \citenamefont {Pauka}, \citenamefont {Nguyen},
  \citenamefont {van Zanten}, \citenamefont {O\ensuremath{'}Farrell},
  \citenamefont {Suter}, \citenamefont {Krogstrup}, \citenamefont {Kuemmeth},\
  and\ \citenamefont {Marcus}}]{DavydasPRA2019}%
  \BibitemOpen
  \bibfield  {author} {\bibinfo {author} {\bibfnamefont {D.}~\bibnamefont
  {Razmadze}}, \bibinfo {author} {\bibfnamefont {D.}~\bibnamefont {Sabonis}},
  \bibinfo {author} {\bibfnamefont {F.~K.}\ \bibnamefont {Malinowski}},
  \bibinfo {author} {\bibfnamefont {G.~C.}\ \bibnamefont {M\'enard}}, \bibinfo
  {author} {\bibfnamefont {S.}~\bibnamefont {Pauka}}, \bibinfo {author}
  {\bibfnamefont {H.}~\bibnamefont {Nguyen}}, \bibinfo {author} {\bibfnamefont
  {D.~M.}\ \bibnamefont {van Zanten}}, \bibinfo {author} {\bibfnamefont
  {E.~C.}\ \bibnamefont {O\ensuremath{'}Farrell}}, \bibinfo {author}
  {\bibfnamefont {J.}~\bibnamefont {Suter}}, \bibinfo {author} {\bibfnamefont
  {P.}~\bibnamefont {Krogstrup}}, \bibinfo {author} {\bibfnamefont
  {F.}~\bibnamefont {Kuemmeth}},\ and\ \bibinfo {author} {\bibfnamefont
  {C.~M.}\ \bibnamefont {Marcus}},\ }\bibfield  {title} {\bibinfo {title}
  {Radio-frequency methods for majorana-based quantum devices: Fast charge
  sensing and phase-diagram mapping},\ }\href
  {https://doi.org/10.1103/PhysRevApplied.11.064011} {\bibfield  {journal}
  {\bibinfo  {journal} {Phys. Rev. Applied}\ }\textbf {\bibinfo {volume}
  {11}},\ \bibinfo {pages} {064011} (\bibinfo {year} {2019})}\BibitemShut
  {NoStop}%
\bibitem [{\citenamefont {Sabonis}\ \emph {et~al.}(2019)\citenamefont
  {Sabonis}, \citenamefont {O'Farrell}, \citenamefont {Razmadze}, \citenamefont
  {van Zanten}, \citenamefont {Suter}, \citenamefont {Krogstrup},\ and\
  \citenamefont {Marcus}}]{Sabonis2019Sep}%
  \BibitemOpen
  \bibfield  {author} {\bibinfo {author} {\bibfnamefont {D.}~\bibnamefont
  {Sabonis}}, \bibinfo {author} {\bibfnamefont {E.~C.~T.}\ \bibnamefont
  {O'Farrell}}, \bibinfo {author} {\bibfnamefont {D.}~\bibnamefont {Razmadze}},
  \bibinfo {author} {\bibfnamefont {D.~M.~T.}\ \bibnamefont {van Zanten}},
  \bibinfo {author} {\bibfnamefont {J.}~\bibnamefont {Suter}}, \bibinfo
  {author} {\bibfnamefont {P.}~\bibnamefont {Krogstrup}},\ and\ \bibinfo
  {author} {\bibfnamefont {C.~M.}\ \bibnamefont {Marcus}},\ }\bibfield  {title}
  {\bibinfo {title} {{Dispersive sensing in hybrid InAs/Al nanowires}},\ }\href
  {https://doi.org/10.1063/1.5116377} {\bibfield  {journal} {\bibinfo
  {journal} {Appl. Phys. Lett.}\ }\textbf {\bibinfo {volume} {115}},\ \bibinfo
  {pages} {102601} (\bibinfo {year} {2019})}\BibitemShut {NoStop}%
\bibitem [{\citenamefont {van Veen}\ \emph {et~al.}(2019)\citenamefont {van
  Veen}, \citenamefont {de~Jong}, \citenamefont {Han}, \citenamefont {Prosko},
  \citenamefont {Krogstrup}, \citenamefont {Watson}, \citenamefont
  {Kouwenhoven},\ and\ \citenamefont {Pfaff}}]{vanVeenPRB2019}%
  \BibitemOpen
  \bibfield  {author} {\bibinfo {author} {\bibfnamefont {J.}~\bibnamefont {van
  Veen}}, \bibinfo {author} {\bibfnamefont {D.}~\bibnamefont {de~Jong}},
  \bibinfo {author} {\bibfnamefont {L.}~\bibnamefont {Han}}, \bibinfo {author}
  {\bibfnamefont {C.}~\bibnamefont {Prosko}}, \bibinfo {author} {\bibfnamefont
  {P.}~\bibnamefont {Krogstrup}}, \bibinfo {author} {\bibfnamefont {J.~D.}\
  \bibnamefont {Watson}}, \bibinfo {author} {\bibfnamefont {L.~P.}\
  \bibnamefont {Kouwenhoven}},\ and\ \bibinfo {author} {\bibfnamefont
  {W.}~\bibnamefont {Pfaff}},\ }\bibfield  {title} {\bibinfo {title} {Revealing
  charge-tunneling processes between a quantum dot and a superconducting island
  through gate sensing},\ }\href {https://doi.org/10.1103/PhysRevB.100.174508}
  {\bibfield  {journal} {\bibinfo  {journal} {Phys. Rev. B}\ }\textbf {\bibinfo
  {volume} {100}},\ \bibinfo {pages} {174508} (\bibinfo {year}
  {2019})}\BibitemShut {NoStop}%
\bibitem [{\citenamefont {van Woerkom}\ \emph {et~al.}(2015)\citenamefont {van
  Woerkom}, \citenamefont {Geresdi},\ and\ \citenamefont
  {Kouwenhoven}}]{vanWoerkom2015Jul}%
  \BibitemOpen
  \bibfield  {author} {\bibinfo {author} {\bibfnamefont {D.~J.}\ \bibnamefont
  {van Woerkom}}, \bibinfo {author} {\bibfnamefont {A.}~\bibnamefont
  {Geresdi}},\ and\ \bibinfo {author} {\bibfnamefont {L.~P.}\ \bibnamefont
  {Kouwenhoven}},\ }\bibfield  {title} {\bibinfo {title} {{One minute parity
  lifetime of a NbTiN Cooper-pair transistor - Nature Physics}},\ }\href
  {https://doi.org/10.1038/nphys3342} {\bibfield  {journal} {\bibinfo
  {journal} {Nat. Phys.}\ }\textbf {\bibinfo {volume} {11}},\ \bibinfo {pages}
  {547} (\bibinfo {year} {2015})}\BibitemShut {NoStop}%
\bibitem [{\citenamefont {Pendharkar}\ \emph {et~al.}(2021)\citenamefont
  {Pendharkar}, \citenamefont {Zhang}, \citenamefont {Wu}, \citenamefont
  {Zarassi}, \citenamefont {Zhang}, \citenamefont {Dempsey}, \citenamefont
  {Lee}, \citenamefont {Harrington}, \citenamefont {Badawy}, \citenamefont
  {Gazibegovic}, \citenamefont {Veld}, \citenamefont {Rossi}, \citenamefont
  {Jung}, \citenamefont {Chen}, \citenamefont {Verheijen}, \citenamefont
  {Hocevar}, \citenamefont {Bakkers}, \citenamefont {Palmstr{\o}m},\ and\
  \citenamefont {Frolov}}]{Pendharkar2021Apr}%
  \BibitemOpen
  \bibfield  {author} {\bibinfo {author} {\bibfnamefont {M.}~\bibnamefont
  {Pendharkar}}, \bibinfo {author} {\bibfnamefont {B.}~\bibnamefont {Zhang}},
  \bibinfo {author} {\bibfnamefont {H.}~\bibnamefont {Wu}}, \bibinfo {author}
  {\bibfnamefont {A.}~\bibnamefont {Zarassi}}, \bibinfo {author} {\bibfnamefont
  {P.}~\bibnamefont {Zhang}}, \bibinfo {author} {\bibfnamefont {C.~P.}\
  \bibnamefont {Dempsey}}, \bibinfo {author} {\bibfnamefont {J.~S.}\
  \bibnamefont {Lee}}, \bibinfo {author} {\bibfnamefont {S.~D.}\ \bibnamefont
  {Harrington}}, \bibinfo {author} {\bibfnamefont {G.}~\bibnamefont {Badawy}},
  \bibinfo {author} {\bibfnamefont {S.}~\bibnamefont {Gazibegovic}}, \bibinfo
  {author} {\bibfnamefont {R.~L. M. O.~h.}\ \bibnamefont {Veld}}, \bibinfo
  {author} {\bibfnamefont {M.}~\bibnamefont {Rossi}}, \bibinfo {author}
  {\bibfnamefont {J.}~\bibnamefont {Jung}}, \bibinfo {author} {\bibfnamefont
  {A.-H.}\ \bibnamefont {Chen}}, \bibinfo {author} {\bibfnamefont {M.~A.}\
  \bibnamefont {Verheijen}}, \bibinfo {author} {\bibfnamefont {M.}~\bibnamefont
  {Hocevar}}, \bibinfo {author} {\bibfnamefont {E.~P. A.~M.}\ \bibnamefont
  {Bakkers}}, \bibinfo {author} {\bibfnamefont {C.~J.}\ \bibnamefont
  {Palmstr{\o}m}},\ and\ \bibinfo {author} {\bibfnamefont {S.~M.}\ \bibnamefont
  {Frolov}},\ }\bibfield  {title} {\bibinfo {title} {{Parity-preserving and
  magnetic field{\textendash}resilient superconductivity in InSb nanowires with
  Sn shells}},\ }\href {https://www.science.org/doi/10.1126/science.aba5211}
  {\bibfield  {journal} {\bibinfo  {journal} {Science}\ }\textbf {\bibinfo
  {volume} {372}},\ \bibinfo {pages} {508} (\bibinfo {year}
  {2021})}\BibitemShut {NoStop}%
\bibitem [{\citenamefont {Kanne}\ \emph
  {et~al.}(2021{\natexlab{b}})\citenamefont {Kanne}, \citenamefont {Marnauza},
  \citenamefont {Olsteins}, \citenamefont {Carrad}, \citenamefont {Sestoft},
  \citenamefont {de~Bruijckere}, \citenamefont {Zeng}, \citenamefont {Johnson},
  \citenamefont {Olsson}, \citenamefont {Grove-Rasmussen},\ and\ \citenamefont
  {Nyg{\aa}rd}}]{Kanne2021Jul}%
  \BibitemOpen
  \bibfield  {author} {\bibinfo {author} {\bibfnamefont {T.}~\bibnamefont
  {Kanne}}, \bibinfo {author} {\bibfnamefont {M.}~\bibnamefont {Marnauza}},
  \bibinfo {author} {\bibfnamefont {D.}~\bibnamefont {Olsteins}}, \bibinfo
  {author} {\bibfnamefont {D.~J.}\ \bibnamefont {Carrad}}, \bibinfo {author}
  {\bibfnamefont {J.~E.}\ \bibnamefont {Sestoft}}, \bibinfo {author}
  {\bibfnamefont {J.}~\bibnamefont {de~Bruijckere}}, \bibinfo {author}
  {\bibfnamefont {L.}~\bibnamefont {Zeng}}, \bibinfo {author} {\bibfnamefont
  {E.}~\bibnamefont {Johnson}}, \bibinfo {author} {\bibfnamefont
  {E.}~\bibnamefont {Olsson}}, \bibinfo {author} {\bibfnamefont
  {K.}~\bibnamefont {Grove-Rasmussen}},\ and\ \bibinfo {author} {\bibfnamefont
  {J.}~\bibnamefont {Nyg{\aa}rd}},\ }\bibfield  {title} {\bibinfo {title}
  {{Epitaxial Pb on InAs nanowires for quantum devices - Nature
  Nanotechnology}},\ }\href {https://doi.org/10.1038/s41565-021-00900-9}
  {\bibfield  {journal} {\bibinfo  {journal} {Nat. Nanotechnol.}\ }\textbf
  {\bibinfo {volume} {16}},\ \bibinfo {pages} {776} (\bibinfo {year}
  {2021}{\natexlab{b}})}\BibitemShut {NoStop}%
\bibitem [{\citenamefont {Bjergfelt}\ \emph {et~al.}(2021)\citenamefont
  {Bjergfelt}, \citenamefont {Carrad}, \citenamefont {Kanne}, \citenamefont
  {Johnson}, \citenamefont {Fiordaliso}, \citenamefont {Jespersen},\ and\
  \citenamefont {Nyg{\aa}rd}}]{Bjergfelt2021Dec}%
  \BibitemOpen
  \bibfield  {author} {\bibinfo {author} {\bibfnamefont {M.~S.}\ \bibnamefont
  {Bjergfelt}}, \bibinfo {author} {\bibfnamefont {D.~J.}\ \bibnamefont
  {Carrad}}, \bibinfo {author} {\bibfnamefont {T.}~\bibnamefont {Kanne}},
  \bibinfo {author} {\bibfnamefont {E.}~\bibnamefont {Johnson}}, \bibinfo
  {author} {\bibfnamefont {E.~M.}\ \bibnamefont {Fiordaliso}}, \bibinfo
  {author} {\bibfnamefont {T.~S.}\ \bibnamefont {Jespersen}},\ and\ \bibinfo
  {author} {\bibfnamefont {J.}~\bibnamefont {Nyg{\aa}rd}},\ }\bibfield  {title}
  {\bibinfo {title} {{Superconductivity and Parity Preservation in As-Grown In
  Islands on InAs Nanowires}},\ }\href
  {https://doi.org/10.1021/acs.nanolett.1c02487} {\bibfield  {journal}
  {\bibinfo  {journal} {Nano Lett.}\ }\textbf {\bibinfo {volume} {21}},\
  \bibinfo {pages} {9875} (\bibinfo {year} {2021})}\BibitemShut {NoStop}%
\bibitem [{\citenamefont {Vaitiek\ifmmode~\dot{e}\else \.{e}\fi{}nas}\ \emph
  {et~al.}(2018)\citenamefont {Vaitiek\ifmmode~\dot{e}\else \.{e}\fi{}nas},
  \citenamefont {Whiticar}, \citenamefont {Deng}, \citenamefont {Krizek},
  \citenamefont {Sestoft}, \citenamefont {Palmstr\o{}m}, \citenamefont
  {Marti-Sanchez}, \citenamefont {Arbiol}, \citenamefont {Krogstrup},
  \citenamefont {Casparis},\ and\ \citenamefont {Marcus}}]{VaitiekenasPRL2018}%
  \BibitemOpen
  \bibfield  {author} {\bibinfo {author} {\bibfnamefont {S.}~\bibnamefont
  {Vaitiek\ifmmode~\dot{e}\else \.{e}\fi{}nas}}, \bibinfo {author}
  {\bibfnamefont {A.~M.}\ \bibnamefont {Whiticar}}, \bibinfo {author}
  {\bibfnamefont {M.-T.}\ \bibnamefont {Deng}}, \bibinfo {author}
  {\bibfnamefont {F.}~\bibnamefont {Krizek}}, \bibinfo {author} {\bibfnamefont
  {J.~E.}\ \bibnamefont {Sestoft}}, \bibinfo {author} {\bibfnamefont {C.~J.}\
  \bibnamefont {Palmstr\o{}m}}, \bibinfo {author} {\bibfnamefont
  {S.}~\bibnamefont {Marti-Sanchez}}, \bibinfo {author} {\bibfnamefont
  {J.}~\bibnamefont {Arbiol}}, \bibinfo {author} {\bibfnamefont
  {P.}~\bibnamefont {Krogstrup}}, \bibinfo {author} {\bibfnamefont
  {L.}~\bibnamefont {Casparis}},\ and\ \bibinfo {author} {\bibfnamefont
  {C.~M.}\ \bibnamefont {Marcus}},\ }\bibfield  {title} {\bibinfo {title}
  {Selective-area-grown semiconductor-superconductor hybrids: A basis for
  topological networks},\ }\href
  {https://doi.org/10.1103/PhysRevLett.121.147701} {\bibfield  {journal}
  {\bibinfo  {journal} {Phys. Rev. Lett.}\ }\textbf {\bibinfo {volume} {121}},\
  \bibinfo {pages} {147701} (\bibinfo {year} {2018})}\BibitemShut {NoStop}%
\bibitem [{\citenamefont {Fano}(1961)}]{Fano}%
  \BibitemOpen
  \bibfield  {author} {\bibinfo {author} {\bibfnamefont {U.}~\bibnamefont
  {Fano}},\ }\bibfield  {title} {\bibinfo {title} {Effects of configuration
  interaction on intensities and phase shifts},\ }\href
  {https://doi.org/10.1103/PhysRev.124.1866} {\bibfield  {journal} {\bibinfo
  {journal} {Phys. Rev.}\ }\textbf {\bibinfo {volume} {124}},\ \bibinfo {pages}
  {1866} (\bibinfo {year} {1961})}\BibitemShut {NoStop}%
\bibitem [{\citenamefont {Babi\ifmmode~\acute{c}\else \'{c}\fi{}}\ and\
  \citenamefont {Sch\"onenberger}(2004)}]{Fanosinglewall}%
  \BibitemOpen
  \bibfield  {author} {\bibinfo {author} {\bibfnamefont {B.}~\bibnamefont
  {Babi\ifmmode~\acute{c}\else \'{c}\fi{}}}\ and\ \bibinfo {author}
  {\bibfnamefont {C.}~\bibnamefont {Sch\"onenberger}},\ }\bibfield  {title}
  {\bibinfo {title} {Observation of fano resonances in single-wall carbon
  nanotubes},\ }\href {https://doi.org/10.1103/PhysRevB.70.195408} {\bibfield
  {journal} {\bibinfo  {journal} {Phys. Rev. B}\ }\textbf {\bibinfo {volume}
  {70}},\ \bibinfo {pages} {195408} (\bibinfo {year} {2004})}\BibitemShut
  {NoStop}%
\bibitem [{\citenamefont {Prada}\ \emph {et~al.}(2020)\citenamefont {Prada},
  \citenamefont {San-Jose}, \citenamefont {de~Moor}, \citenamefont {Geresdi},
  \citenamefont {Lee}, \citenamefont {Klinovaja}, \citenamefont {Loss},
  \citenamefont {Nyg{\aa}rd}, \citenamefont {Aguado},\ and\ \citenamefont
  {Kouwenhoven}}]{Prada2020Oct}%
  \BibitemOpen
  \bibfield  {author} {\bibinfo {author} {\bibfnamefont {E.}~\bibnamefont
  {Prada}}, \bibinfo {author} {\bibfnamefont {P.}~\bibnamefont {San-Jose}},
  \bibinfo {author} {\bibfnamefont {M.~W.~A.}\ \bibnamefont {de~Moor}},
  \bibinfo {author} {\bibfnamefont {A.}~\bibnamefont {Geresdi}}, \bibinfo
  {author} {\bibfnamefont {E.~J.~H.}\ \bibnamefont {Lee}}, \bibinfo {author}
  {\bibfnamefont {J.}~\bibnamefont {Klinovaja}}, \bibinfo {author}
  {\bibfnamefont {D.}~\bibnamefont {Loss}}, \bibinfo {author} {\bibfnamefont
  {J.}~\bibnamefont {Nyg{\aa}rd}}, \bibinfo {author} {\bibfnamefont
  {R.}~\bibnamefont {Aguado}},\ and\ \bibinfo {author} {\bibfnamefont {L.~P.}\
  \bibnamefont {Kouwenhoven}},\ }\bibfield  {title} {\bibinfo {title} {{From
  Andreev to Majorana bound states in hybrid
  superconductor{\textendash}semiconductor nanowires}},\ }\href
  {https://doi.org/10.1038/s42254-020-0228-y} {\bibfield  {journal} {\bibinfo
  {journal} {Nat. Rev. Phys.}\ }\textbf {\bibinfo {volume} {2}},\ \bibinfo
  {pages} {575} (\bibinfo {year} {2020})}\BibitemShut {NoStop}%
\bibitem [{\citenamefont {Chiu}\ \emph {et~al.}(2017)\citenamefont {Chiu},
  \citenamefont {Sau},\ and\ \citenamefont {Das~Sarma}}]{Chiu2017}%
  \BibitemOpen
  \bibfield  {author} {\bibinfo {author} {\bibfnamefont {C.-K.}\ \bibnamefont
  {Chiu}}, \bibinfo {author} {\bibfnamefont {J.~D.}\ \bibnamefont {Sau}},\ and\
  \bibinfo {author} {\bibfnamefont {S.}~\bibnamefont {Das~Sarma}},\ }\bibfield
  {title} {\bibinfo {title} {Conductance of a superconducting coulomb-blockaded
  majorana nanowire},\ }\href {https://doi.org/10.1103/PhysRevB.96.054504}
  {\bibfield  {journal} {\bibinfo  {journal} {Phys. Rev. B}\ }\textbf {\bibinfo
  {volume} {96}},\ \bibinfo {pages} {054504} (\bibinfo {year}
  {2017})}\BibitemShut {NoStop}%
\bibitem [{\citenamefont {Lai}\ \emph {et~al.}(2021)\citenamefont {Lai},
  \citenamefont {Das~Sarma},\ and\ \citenamefont {Sau}}]{Yu-HuaPRB2021}%
  \BibitemOpen
  \bibfield  {author} {\bibinfo {author} {\bibfnamefont {Y.-H.}\ \bibnamefont
  {Lai}}, \bibinfo {author} {\bibfnamefont {S.}~\bibnamefont {Das~Sarma}},\
  and\ \bibinfo {author} {\bibfnamefont {J.~D.}\ \bibnamefont {Sau}},\
  }\bibfield  {title} {\bibinfo {title} {Theory of coulomb blockaded transport
  in realistic majorana nanowires},\ }\href
  {https://doi.org/10.1103/PhysRevB.104.085403} {\bibfield  {journal} {\bibinfo
   {journal} {Phys. Rev. B}\ }\textbf {\bibinfo {volume} {104}},\ \bibinfo
  {pages} {085403} (\bibinfo {year} {2021})}\BibitemShut {NoStop}%
\bibitem [{\citenamefont {Jellinggaard}\ \emph {et~al.}(2016)\citenamefont
  {Jellinggaard}, \citenamefont {Grove-Rasmussen}, \citenamefont {Madsen},\
  and\ \citenamefont {Nyg{\aa}rd}}]{Jellinggaard2016Aug}%
  \BibitemOpen
  \bibfield  {author} {\bibinfo {author} {\bibfnamefont {A.}~\bibnamefont
  {Jellinggaard}}, \bibinfo {author} {\bibfnamefont {K.}~\bibnamefont
  {Grove-Rasmussen}}, \bibinfo {author} {\bibfnamefont {M.~H.}\ \bibnamefont
  {Madsen}},\ and\ \bibinfo {author} {\bibfnamefont {J.}~\bibnamefont
  {Nyg{\aa}rd}},\ }\bibfield  {title} {\bibinfo {title} {{Tuning
  Yu-Shiba-Rusinov states in a quantum dot}},\ }\href
  {https://doi.org/10.1103/PhysRevB.94.064520} {\bibfield  {journal} {\bibinfo
  {journal} {Phys. Rev. B}\ }\textbf {\bibinfo {volume} {94}},\ \bibinfo
  {pages} {064520} (\bibinfo {year} {2016})}\BibitemShut {NoStop}%
\bibitem [{\citenamefont {Mikkelsen}\ \emph {et~al.}(2018)\citenamefont
  {Mikkelsen}, \citenamefont {Kotetes}, \citenamefont {Krogstrup},\ and\
  \citenamefont {Flensberg}}]{PhysRevX.8.031040}%
  \BibitemOpen
  \bibfield  {author} {\bibinfo {author} {\bibfnamefont {A.~E.~G.}\
  \bibnamefont {Mikkelsen}}, \bibinfo {author} {\bibfnamefont {P.}~\bibnamefont
  {Kotetes}}, \bibinfo {author} {\bibfnamefont {P.}~\bibnamefont {Krogstrup}},\
  and\ \bibinfo {author} {\bibfnamefont {K.}~\bibnamefont {Flensberg}},\
  }\bibfield  {title} {\bibinfo {title} {Hybridization at
  superconductor-semiconductor interfaces},\ }\href
  {https://doi.org/10.1103/PhysRevX.8.031040} {\bibfield  {journal} {\bibinfo
  {journal} {Phys. Rev. X}\ }\textbf {\bibinfo {volume} {8}},\ \bibinfo {pages}
  {031040} (\bibinfo {year} {2018})}\BibitemShut {NoStop}%
\bibitem [{\citenamefont {Iftikhar}\ \emph {et~al.}(2015)\citenamefont
  {Iftikhar}, \citenamefont {Jezouin}, \citenamefont {Anthore}, \citenamefont
  {Gennser}, \citenamefont {Parmentier}, \citenamefont {Cavanna},\ and\
  \citenamefont {Pierre}}]{IftikharNat2015}%
  \BibitemOpen
  \bibfield  {author} {\bibinfo {author} {\bibfnamefont {Z.}~\bibnamefont
  {Iftikhar}}, \bibinfo {author} {\bibfnamefont {S.}~\bibnamefont {Jezouin}},
  \bibinfo {author} {\bibfnamefont {A.}~\bibnamefont {Anthore}}, \bibinfo
  {author} {\bibfnamefont {U.}~\bibnamefont {Gennser}}, \bibinfo {author}
  {\bibfnamefont {F.~D.}\ \bibnamefont {Parmentier}}, \bibinfo {author}
  {\bibfnamefont {A.}~\bibnamefont {Cavanna}},\ and\ \bibinfo {author}
  {\bibfnamefont {F.}~\bibnamefont {Pierre}},\ }\bibfield  {title} {\bibinfo
  {title} {Two-channel kondo effect and renormalization flow with macroscopic
  quantum charge states},\ }\href {https://doi.org/10.1038/nature15384}
  {\bibfield  {journal} {\bibinfo  {journal} {Nature}\ }\textbf {\bibinfo
  {volume} {526}},\ \bibinfo {pages} {233} (\bibinfo {year}
  {2015})}\BibitemShut {NoStop}%
\bibitem [{\citenamefont {Iftikhar}\ \emph {et~al.}(2018)\citenamefont
  {Iftikhar}, \citenamefont {Anthore}, \citenamefont {Mitchell}, \citenamefont
  {Parmentier}, \citenamefont {Gennser}, \citenamefont {Ouerghi}, \citenamefont
  {Cavanna}, \citenamefont {Mora}, \citenamefont {Simon},\ and\ \citenamefont
  {Pierre}}]{IftikharScience2018}%
  \BibitemOpen
  \bibfield  {author} {\bibinfo {author} {\bibfnamefont {Z.}~\bibnamefont
  {Iftikhar}}, \bibinfo {author} {\bibfnamefont {A.}~\bibnamefont {Anthore}},
  \bibinfo {author} {\bibfnamefont {A.~K.}\ \bibnamefont {Mitchell}}, \bibinfo
  {author} {\bibfnamefont {F.~D.}\ \bibnamefont {Parmentier}}, \bibinfo
  {author} {\bibfnamefont {U.}~\bibnamefont {Gennser}}, \bibinfo {author}
  {\bibfnamefont {A.}~\bibnamefont {Ouerghi}}, \bibinfo {author} {\bibfnamefont
  {A.}~\bibnamefont {Cavanna}}, \bibinfo {author} {\bibfnamefont
  {C.}~\bibnamefont {Mora}}, \bibinfo {author} {\bibfnamefont {P.}~\bibnamefont
  {Simon}},\ and\ \bibinfo {author} {\bibfnamefont {F.}~\bibnamefont
  {Pierre}},\ }\bibfield  {title} {\bibinfo {title} {Tunable quantum
  criticality and super-ballistic transport in a charge kondo circuit},\ }\href
  {https://doi.org/10.1126/science.aan5592} {\bibfield  {journal} {\bibinfo
  {journal} {Science}\ }\textbf {\bibinfo {volume} {360}},\ \bibinfo {pages}
  {1315} (\bibinfo {year} {2018})}\BibitemShut {NoStop}%
\bibitem [{\citenamefont {Pouse}\ \emph {et~al.}(2021)\citenamefont {Pouse},
  \citenamefont {Peeters}, \citenamefont {Hsueh}, \citenamefont {Gennser},
  \citenamefont {Cavanna}, \citenamefont {Kastner}, \citenamefont {Mitchell},\
  and\ \citenamefont {Goldhaber-Gordon}}]{pousearxiv2021}%
  \BibitemOpen
  \bibfield  {author} {\bibinfo {author} {\bibfnamefont {W.}~\bibnamefont
  {Pouse}}, \bibinfo {author} {\bibfnamefont {L.}~\bibnamefont {Peeters}},
  \bibinfo {author} {\bibfnamefont {C.~L.}\ \bibnamefont {Hsueh}}, \bibinfo
  {author} {\bibfnamefont {U.}~\bibnamefont {Gennser}}, \bibinfo {author}
  {\bibfnamefont {A.}~\bibnamefont {Cavanna}}, \bibinfo {author} {\bibfnamefont
  {M.~A.}\ \bibnamefont {Kastner}}, \bibinfo {author} {\bibfnamefont {A.~K.}\
  \bibnamefont {Mitchell}},\ and\ \bibinfo {author} {\bibfnamefont
  {D.}~\bibnamefont {Goldhaber-Gordon}},\ }\href@noop {} {\bibinfo {title}
  {Exotic quantum critical point in a two-site charge kondo circuit}} (\bibinfo
  {year} {2021}),\ \Eprint {https://arxiv.org/abs/2108.12691} {arXiv:2108.12691
  [cond-mat.mes-hall]} \BibitemShut {NoStop}%
\end{thebibliography}%

\end{document}